\documentclass[preprint2]{aastex}

\shorttitle{URAT SOUTH PARALLAX RESULTS: DISCOVERY OF NEW NEARBY STARS}
\shortauthors{Finch \& Zacharias}

\def\arcsec{\hbox{$^{\prime\prime}$}}
\def\farcs{\hbox{$.\!\!^{\prime\prime}$}}

\begin{document}

\title{URAT SOUTH PARALLAX RESULTS}
\author{Charlie T. Finch, Norbert Zacharias}

\email{charlie.finch@navy.mil}

\affil{U.S. Naval Observatory, Washington DC 20392--5420}

\author{Wei-Chun Jao}

\affil{Georgia State University, Atlanta, GA 30302--4106}

\author{}
\affil{}

%%%%%%%%%%%%%%%%%%%%%%%%%%%%%%%%%%%%%%%%%%%%%%%%%%%%%%%%%%%%%%%%%%%%%%%%%%%%%%
% {Abstract}
%%%%%%%%%%%%%%%%%%%%%%%%%%%%%%%%%%%%%%%%%%%%%%%%%%%%%%%%%%%%%%%%%%%%%%%%%%%%%%

\begin{abstract}

We present 916 trigonometric parallaxes and proper motions of newly
discovered nearby stars from the United States Naval Observatory
(USNO) Robotic Astrometric Telescope (URAT). Observations were taken
at the Cerro Tololo Interamerican Observatory (CTIO) over a 2 year
period from Oct 2015 to Oct 2017 covering the entire sky south of
about $+25^{\circ}$ declination.  SPM4 and UCAC4 early epoch catalog
data were added to extend the temporal coverage for the parallax and
proper motion fit up to 48 years. Using these new URAT parallaxes,
optical and near-IR photometry from the APASS and 2MASS catalogs, we
identify possible new nearby dwarfs, young stars, low-metallicity
subdwarfs and white dwarfs. Comparison to known trigonometric
parallaxes show a high quality of the URAT-based results confirming
the error in parallax of the URAT south parallaxes reported here to be
between 2 and 13 mas. We also include additional 729 trigonometric
parallaxes from the URAT north 25 pc sample published in \cite{upc25}
here after applying the same criterion as for the southern sample to
have a complete URAT 25 pc sample presented in this paper.

% mean error of upcs_25_v4 list = 5.0
% mean error of upcn_25_v4 list = 6.1
% mean error of upcs_known_new  =

\end{abstract}

\keywords{parallaxes --- solar neighborhood --- stars: distances --- stars:
statistics --- surveys --- astrometry --- photometry --- double stars: CPM} 

%%%%%%%%%%%%%%%%%%%%%%%%%%%%%%%%%%%%%%%%%%%%%%%%%%%%%%%%%%%%%%%%%%%%%%%%%%%%%%
\section{INTRODUCTION}
\label{sec:introduction}
%%%%%%%%%%%%%%%%%%%%%%%%%%%%%%%%%%%%%%%%%%%%%%%%%%%%%%%%%%%%%%%%%%%%%%%%%%%%%%

Nearby stars are important in the investigation of stellar properties
due to their proximity to the sun.  Accurate distances are important
for investigating the multiplicity of stellar systems, luminosity,
masses, sizes and ages of objects, and exoplanet research.  By using
the method of trigonometric parallax we are able to directly measure
accurate stellar distances requiring only the Earth's orbital
motion and multiple epoch astrometric observations of the target star.

This study is a continuation of our short-term ground-based effort to
obtain trigonometric parallaxes with the United States Naval
Observatory (USNO) Robotic Astrometric Telescope (URAT) \citep{urat1}
without prior selection of target stars by e.g.~high proper motion or
photometric indication prior to Gaia.  Our long term goal for our
Southern observing program remains to focus on bright star observing
with a catalog to be published.

Results from our northern observations were
published as URAT Parallax Catalog (UPC) \citep{upc}.  We present also
in this paper our URAT north 25 pc sample utilizing the same stringent
cuts used for the URAT south data to have a complete URAT 25 pc
sample in one place.  
Our southern hemisphere observing program is still ongoing, now
concentrating on the very bright stars to supplement the upcoming
European Space Agency (ESA) Gaia mission second data release (DR2)
expected in April 2018, which will include a more comprehensive
inventory of nearby stars extending much further from the Sun as this
study can accomplish.

Here we identify 1526 stars within 25 pc of the Sun and south of
$+25^{\circ}$ declination of which 916 have no previously published
trigonometric parallax.  We find 64 stars within 10 pc with 5 having
no previously published trigonometric parallax.  Visual inspection of
each parallax fit diagnostic plot page (path on sky, distribution of 
residuals, parallax factor and epoch coverage)
was performed for all stars discovered in this investigation.  
In addition, a visual inspection of Digitized Sky
Survey (DSS) data was performed for all stars found to be within 10 pc
or having odd placements on the Hertzsprung-Russell Diagram (HRD) and
Color-Color diagram.

Comparison of our trigonometric parallax results with other known
trigonometric parallax data are performed for sources in common and no
significant biases were found confirming the error estimate on our
parallax to be on the 2 to 13 mas level depending on observing
history and brightness of the target.  Our trigonometric parallax
results are published in the URAT Parallax Catalog South (UPCs),
however, no public release of the entire URAT south data set including
observed positions for about 300 million stars is planned at this
time.

%%%%%%%%%%%%%%%%%%%%%%%%%%%%%%%%%%%%%%%%%%%%%%%%%%%%%%%%%%%%%%%%%%%%%%%%%%%%%%
\section {OBSERVATIONS}
\label{sec:observing}
%%%%%%%%%%%%%%%%%%%%%%%%%%%%%%%%%%%%%%%%%%%%%%%%%%%%%%%%%%%%%%%%%%%%%%%%%%%%%%

URAT observations were performed at the Cerro Tololo Interamerican
Observatory (CTIO) similar to those in the north \citep{upc25},
however, with 3 major differences: 1) All observations in the south
were made with a 4.5-mag attenuation objective grating in front of the
lens.  2) Instead of the 60 and 240 sec exposures on a regular survey
field, 4 exposures were taken with 60, 30, 10 and another 10 sec.  3)
In addition to these all-sky regular survey observations, stars of
URAT bandpass magnitude 4.5 and brighter were observed individually
through a neutral density (ND) spot filter of about 4.5 mag
attenuation which is large enough to cover also the 1st order grating
images.  These changes from the north survey have been made to
accommodate the main focus of Southern survey, bright star observing.

Thus, the southern URAT observations do not go as deep as those in the
north (limiting URAT magnitude of about 17.5 vs.~18.5).  However, this
scheme allows access to even the brightest stars up to Sirius.  The ND
spot observations of the very bright stars use 5 to 60 sec exposures,
depending on brightness.  Results presented here solely rely on the
central images of the general survey (i.e.~not ND spot data) with a
saturation limit of about URAT magnitude 8.  URAT results on the
brightest stars will be published at a later time.

URAT observes through a single filter which is part of the dewar
window to provide a fixed bandpass (about 680 to 760 nm).  All URAT
south observations have been taken close to the meridian, typically
with an hour angle within $\pm 5^{\circ}$.  For this survey as in the
north, the observing nights are split into 5 equal long periods during
which a different set of 3 dither positions of a field are observed.
This gives us a sufficient parallactic angle distribution over the
observing year.  Sky coverage color-coded with number of 10-second
exposures taken on individual 0.25 by 0.25 degree patches of the sky
is presented in Figure~\ref{skyurats}.  Note the pattern of the 4 CCDs
URAT focal plane is seen around a few very bright stars which were
observed through the ND filter with 10 sec exposures, while all the
rest of the data are from the regular survey.

The URAT telescope has a clear aperture of 206 mm with a 2 m focal
length.  With 4 large CCDs in the focal plane a single exposure covers
28 square degrees with a resolution of 0.9 arcsecond/pixel.  Each CCD
in the focal plane covers a 2.65 by 2.65 degree area on the sky.  Data
spanning $\approx$ 2 years of operations (October 2015 to October 2017)
are used for this parallax investigation.  For more details about the
URAT program and instrument we refer the reader to the URAT1 paper
\citep{urat1}.

In order to better separate parallax and proper motion, early epoch
data were added to the URAT observations.  Mean positions at mean
epoch were extracted from both the fourth USNO CCD Astrograph Catalog
(UCAC4) \citep{ucac4} and the fourth Southern Proper Motion catalog
(SPM4) \citep{spm4}.  Note, UCAC4 is a compiled catalog which includes
many more catalogs than the UCAC observations taken between 1997 and
2004, thus many stars in our sample have a much earlier epoch from
UCAC4.  UCAC4 covers the entire sky with a limiting magnitude of about
R = 16.5 mag, while SPM4 with observations taken between 1965 and 2008
goes deeper, covering the entire URAT magnitude range but is limited
to the sky area south of $-20^{\circ}$ deg declination.  The
distribution of total epoch span of stars in our investigation is
shown in Figure~\ref{tepoch}.  This has been cut at 28 years to help
better show the separation gap between URAT data matched and
without any match to the UCAC4 or SPM4 early data.

%%%%%%%%%%%%%%%%%%%%%%%%%%%%%%%%%%%%%%%%%%%%%%%%%%%%%%%%%%%%%%%%%%%%%%%%%%%%%%
\section {ASTROMETRIC REDUCTIONS}
\label{sec:astrometricreductions}
%%%%%%%%%%%%%%%%%%%%%%%%%%%%%%%%%%%%%%%%%%%%%%%%%%%%%%%%%%%%%%%%%%%%%%%%%%%%%%

%%%%%%%%%%%%%%%%%%%%%%%%%%%%%%%%%%%%%%%%%%%%%%%%%%%%%%%%%%%%%%%%%%%%%%%%%%%%%%
\subsection {Raw Data Processing}
\label{sec:rawdataprocessing}
%%%%%%%%%%%%%%%%%%%%%%%%%%%%%%%%%%%%%%%%%%%%%%%%%%%%%%%%%%%%%%%%%%%%%%%%%%%%%%

All raw and processed images have been bias corrected with dark and
flat-field corrections applied to the 2-byte-integer FITS files using
custom code.  We use the same methods as in \citep{upc25} for detecting
stellar images with a 4-sigma threshold above the background.  Object
centers have been determined using custom code to perform
2-dimensional spherical symmetric Gaussian model profile fits of the
observed stellar images using the processed pixel data.  A significant
contribution to the observed point spread function (PSF) comes from
seeing and also diffraction due to the small aperture which leads to
an observed image profile width of about 2 pixels full width at half
maximum (FWHM).  However, the PSF is uniform across the entire field
of view allowing us to use one model function across the entire focal
plane.

Grating images of order 1 and 2 were identified with custom code based
on the expected brightness, location with respect to a central image
and image elongation.  Higher order images are too elongated to make
the list of acceptable images in the object detection code.
A mean position from the 1st order grating images is propagated to the
astrometric reductions together with all fitted central image
positions (all in pixel coordinate space).  Then all individually
detected grating images are removed from the observed
objects list, before the RA,Dec is calculated.

Although great care has been taken to make this step as accurate as
possible, some spurious detection's associated with grating images
possibly, impacted by blended images will have made it into the
astrometric reduction step.  However, the parallax results presented
here should be free of those contaminations because of conservative
cuts applied to the data throughout this investigation and outlier
rejections applied at various stages of the reduction process.  Such
``left-over" grating image contamination would be much more of a
problem for a general URAT south catalog based on all available data.

%%%%%%%%%%%%%%%%%%%%%%%%%%%%%%%%%%%%%%%%%%%%%%%%%%%%%%%%%%%%%%%%%%%%%%%%%%%%%%
\subsection {Reference Star Catalog}
\label{sec:referencestarcatalog}
%%%%%%%%%%%%%%%%%%%%%%%%%%%%%%%%%%%%%%%%%%%%%%%%%%%%%%%%%%%%%%%%%%%%%%%%%%%%%%

A special reference star catalog was constructed for this project, the
UCAC5 \citep{ucac5}, instead of using the UCAC4 as we did earlier for
the URAT north data.  Gaia DR1 data \citep{gaia} was used as the basis
of this new reference star catalog providing accurate positions of
stars in the about 5 to 21 G-band mag range at epoch 2015.0, close to
our URAT south data observing epoch.  The Tycho Gaia Astrometric
Solution (TGAS) \citep{gaia} proper motions of DR1 were adopted for
those just over 2 million stars. New proper motions were derived for
about 100 million stars using UCAC4 early epoch data and Gaia DR1
positions for stars in common.  Typical errors in proper motions are
between 1 and 5 mas/yr depending on brightness of the star.  Thus
UCAC5 extends the TGAS data with similar precision in proper motions
to about 14th magnitude and with somewhat lower precision to 16th
magnitude.

%%%%%%%%%%%%%%%%%%%%%%%%%%%%%%%%%%%%%%%%%%%%%%%%%%%%%%%%%%%%%%%%%%%%%%%%%%%%%%
\subsection {Astrometric Solution}
\label{sec:astrometricsolution}
%%%%%%%%%%%%%%%%%%%%%%%%%%%%%%%%%%%%%%%%%%%%%%%%%%%%%%%%%%%%%%%%%%%%%%%%%%%%%%

An 8-parameter 'plate' model, the same used in the north has also been
used here for the astrometric reductions (linear + tilt terms).
However, For this investigation we use the UCAC5 (see above) reference
star catalog, restricted to a URAT magnitude of  8 to 15.

We use a conventional weighted least-squares adjustment with outlier
rejection individually on each CCD exposure with typically several
hundred to many thousand reference stars per astrometric solution
(Figure~\ref{asts}).  The weights are calculated based on the total,
formal errors of individual observations. Exposures with less than 100
reference stars on a CCD were rejected as were exposures failing the
observing quality control standards (see URAT1 paper \citep{urat1}).
Using look-up tables from the preliminary reductions and residual
analysis we have corrected the data for both geometric field
distortions (10 to 60 mas) and pixel phase errors (0 to 15 mas).

The URAT data was split into groups of observations spanning about a month
each.  Preliminary astrometric solutions were obtained for each group and
the residuals analyzed.  Small (few mas) positional corrections as a function 
of magnitude were applied to central images, while the grating images
required much larger corrections (up to about 80 mas).
Astrometric reductions were iteratively repeated with updated correction
models.
The distribution of the astrometric solution errors (in unit of mas and
chi-square unit weight) are shown in Figure~\ref{astsdist}.

Epoch positions ($\alpha, \delta$) of all stars have been obtained on
the International Celestial Reference System (ICRS) via UCAC5.  The
errors of individual observations are typically 10 - 60 mas depending
on brightness of the object and exposure time.  This matches our
typical errors in the north due to the shorter exposure times and
signal to noise ratio.  Individual epoch positions which have been
matched to individual stars with mean data and indexing are stored in
a separate large file allowing fast, direct access to the individual
observations.  These URAT data consist of about 400 million individual
objects with about 11.3 billion observations. There are 254 million
stars with 3 or more observations each.  For more details about the
instrument and astrometric reductions, the reader is referred to the
URAT1 paper \citep{urat1}.

%%%%%%%%%%%%%%%%%%%%%%%%%%%%%%%%%%%%%%%%%%%%%%%%%%%%%%%%%%%%%%%%%%%%%%%%%%%%%%
\subsection {Selection and Match with Early Epoch Data}
\label{sec:selection}
%%%%%%%%%%%%%%%%%%%%%%%%%%%%%%%%%%%%%%%%%%%%%%%%%%%%%%%%%%%%%%%%%%%%%%%%%%%%%%

A sub-set of about 155 million stars were extracted from the
astrometric solution mean position data retaining only those stars
which have at least 10 observations and an epoch span of at least 0.9
year (URAT observing).  This set of stars was matched with the UCAC4
and SPM4 catalogs after updating the UCAC4 and SPM4 positions to the
mean URAT south observation epoch of 2016.8 using the UCAC4 and SPM4
proper motions, respectively.

The position comparison of the URAT sub-set with UCAC4 revealed 83.4 million
unique matches within 2 arcsec per coordinate.  Similarly the match with SPM4
gave 83.3 million uniquely matched objects.
The URAT data as well as the UCAC4 and SPM4 matches were labeled with a common,
running ID number which allows easy retrieval of those data as well as the
associated individual URAT observations of each star.

%%%%%%%%%%%%%%%%%%%%%%%%%%%%%%%%%%%%%%%%%%%%%%%%%%%%%%%%%%%%%%%%%%%%%%%%%%%%%%
\subsection {Solving For Parallax}
%%%%%%%%%%%%%%%%%%%%%%%%%%%%%%%%%%%%%%%%%%%%%%%%%%%%%%%%%%%%%%%%%%%%%%%%%%%%%%

The same pipeline as for our northern hemisphere investigation
\citep{upc25} was applied here utilizing routines from
\citep{jaothesis}, the JPL DE405 ephemeris and making use of the
parallax factor \citep{green} for determining parallaxes.  We use each
URAT mid exposure to determine the location of the Earth.  These
rectangular coordinates $X$, $Y$, and $Z$ at epoch are then used to
calculate the parallax factors using the same formulae as our northern
hemisphere investigation shown here:

\begin{equation}
P_{\alpha} \ =  \ X \sin \alpha \ - \ Y \cos \alpha 
\end{equation}
\begin{equation}
P_{\delta} \ = \ X \cos \alpha \sin \delta \ + \ Y \sin \alpha \sin \delta
 \ - \ Z \cos \delta 
\end{equation}

We then use each individual parallax factor corresponding to all
individual data of a given target to simultaneously solve for each
astrometric parameter (mean position, proper motion and parallax)
using only 'good' epoch data in a weighted least-squares adjustment
with outlier rejection from the equations:

\begin{equation}
x(t) \ = \ x(t_{0}) \ + \ \mu_{x} (t - t_{0}) \ + \ \pi P_{\alpha}
\end{equation}
\begin{equation}
y(t) \ = \ y(t_{0}) \ + \ \mu_{y} (t - t_{0}) \ + \ \pi P_{\delta}
\end{equation}

Here as in \citep{upc25} the $x(t), y(t)$ are the positions
of a given star on the tangential plane as function of time ($t$),
$\pi$ is the parallax, $\mu_{x} = \mu_{\alpha} cos\delta$ and 
$\mu_{y}= \mu_{\delta}$ represent the proper motions in RA and DEC,
respectively.  Here we choose the initial instant of time ($t_{0}$ to
be the first observing epoch as our zero point for $x, y$ and $t$.

We show in Table~\ref{table_cuts1} our initial adopted cuts to the
URAT south epoch data of each individual star when solving for
parallax.  These limits have been imposed to not allow saturated
stars, stars with too few photons and stars with poorly determined
positions to be used in the fits.  We impose a number and epoch span
of observation cut as well empirically after comparing the URAT
parallaxes to the Hipparcos new reduction \citep{HIP2} and TGAS
catalogs.

As in \citep{upc25} we add a 10 mas error floor to the random errors before
calculating weights for individual observations.  These weights which
were used in the least-square adjustments when solving for parallax
can vary largely due to the exposure time and amplitude of individual
observations.

We imposed the same 3-sigma outlier rejection criteria as for the
northern data where the largest residuals (typically a few percent)
have been iteratively removed from the parallax fit solution of
individual stars.

%%%%%%%%%%%%%%%%%%%%%%%%%%%%%%%%%%%%%%%%%%%%%%%%%%%%%%%%%%%%%%%%%%%%%%%%%%%%%%
\subsection {Conversion From Relative To Absolute Parallax}
%%%%%%%%%%%%%%%%%%%%%%%%%%%%%%%%%%%%%%%%%%%%%%%%%%%%%%%%%%%%%%%%%%%%%%%%%%%%%%

We use the same method here as in \citep{upc25} to convert the relative
parallaxes from the fit solutions to absolute parallaxes using
photometric parallaxes for the same set of reference stars used in the
reductions of the URAT positions. This method was used as opposed to
the more reliable spectroscopic parallax method due to the lack of
spectroscopic data for millions of stars in our survey.

We run each individual reference star in a given URAT frame through 16
photometric color-$M_{K{_s}}$ relations \citep{ucac4s}.  These
relations make use of the AAVSO Photometric All-Sky Survey (APASS)
$BVgri$ and Two Micron All Sky Survey (2MASS) $JHK_s$ photometry which
are attached to the UCAC4 catalog.  We typically have many hundred to
several thousands of reference stars in the 2 by 2 square degree area
of sky surrounding each target star.  We use this data to obtain a mean
absolute parallax correction for each target star with the assumption
that all stars are main-sequence, due to the lack of information for
each individual star.

The mean parallax correction for each target star in this
investigation is 1.3 mas which varies from 0.7 to 8.4 mas depending on
the field.  Fields with corrections larger than 5.0 mas tend to have
fewer reference stars and larger errors on the corrections. This is in
part due to the limited color range of our photometric
color-$M_{K{_s}}$ relations. For all stars having an unknown
correction, we use the mean absolute correction. An unknown correction
here is possible due to lack of photometry information in a given
field and/or the strict limit of the color range for the relations.
If a star has a correction that is greater than 3 times the average
($\ge$ 3.9 mas), we adopt this as a cutoff and use the 3.9 mas for the
conversion to absolute parallax.  The original parallax correction is
left in the tables for the readers to use as a flag to determine if an
average was used.

\subsection{Biases}

We take the same approach as in \citep{upc25} and do not apply any
corrections to our parallaxes for the Lutz-Kelker bias \citep{LK73}
because we do not draw any conclusions about completeness of a
distance limited sample or interpret the results regarding absolute
luminosity.  The goal of this paper is to present the observed
trigonometric parallaxes for a large number of stars.

We do however concede that our parallax results could be affected by a
statistical bias due to the large number of stars.  As explained in
\citep{upc25} if some observational errors of a distant star with small
parallax happen to follow a much larger parallax path on the sky then
a ``significant parallax'' with acceptable error will come out of the
solution fit.

The chance of this happening will drastically decrease with increased 
epoch span (i.e. observations spanning 5 to 10 years are taken).
However, for our stars we do have only 1 to 2 years epoch span 
plus 1 or 2 observations at an earlier epoch for most stars.
A single group of few observations accidentally offset a certain way
can bias the result significantly.
With a high number of stars with formal parallax solution like in this
investigation, there will be a fraction of "fake" parallaxes just due 
to this statistical bias.  

%%%%%%%%%%%%%%%%%%%%%%%%%%%%%%%%%%%%%%%%%%%%%%%%%%%%%%%%%%%%%%%%%%%%%%%%%%%%%%
\section {RESULTS}
%%%%%%%%%%%%%%%%%%%%%%%%%%%%%%%%%%%%%%%%%%%%%%%%%%%%%%%%%%%%%%%%%%%%%%%%%%%%%%

After running the URAT parallax pipeline over the entire set of URAT
south epoch data we obtain 24.7 million parallax solutions.  We then
compared these results with the Hipparcos new reduction catalog, TGAS,
SIMBAD database \citep{simbad}, REsearch Consortium On Nearby Stars
(RECONS) 25 pc database \citep{RECONS25} and other published
trigonometric parallax data using a search radius up to 60 arcsec to
flag stars with previously published trigonometric parallaxes.  The
remaining results for nearby stars having no prior published
trigonometric parallax have been pulled from this set of data.  Due to
the large distance associated with most stars and possible large
random observational errors most of these millions of stars at this
point have an insignificant parallax solution.

%%%%%%%%%%%%%%%%%%%%%%%%%%%%%%%%%%%%%%%%%%%%%%%%%%%%%%%%%%%%%%%%%%%%%%%%%%%%%%
\subsection {Comparison To Published Parallaxes}
\label{sec:comparison}
%%%%%%%%%%%%%%%%%%%%%%%%%%%%%%%%%%%%%%%%%%%%%%%%%%%%%%%%%%%%%%%%%%%%%%%%%%%%%%

The goal of this paper is to discover new nearby stars within 25 pc.
To determine how well we can detect nearby stars within 25 pc using
our URAT south data we use the Hipparcos new reduction \citep{HIP2}
and TGAS \citep{gaia} catalogs using only the initial cuts from the
URAT parallax pipeline summarized in Table~\ref{table_cuts1}.

The Hipparcos new reduction contains 514 stars with a parallax $\ge$
40 mas, south of declination $+20^{\circ}$ where we have full coverage
and with a Hipparcos magnitude between 9.0 and 16.0 which closely
matches the URAT magnitude range.  URAT recovers 469 (=91.2\%) of this
Hipparcos 25 pc sample.  Most of the stars not recovered are likely
due to the initial amplitude (S/N ratio), parallax solution error and
number of observation cuts which are summarized in
Table~\ref{table_cuts1}.  Of the 469 stars recovered, 417 (=88.9\%)
have a URAT parallax $\ge$ 30 mas.  We show a comparison of the
Hipparcos 25 pc sample with the URAT results in Figure~\ref{th25urat},
where the center line indicates perfect agreement and the 2 outer
lines indicate a 10 mas difference from the Hipparcos parallax values.

The TGAS catalog contains 297 stars with a parallax $\ge$ 40 mas south
of $+20^{\circ}$ declination where URAT has full coverage having a
Gaia magnitude between 9.0 and 16.0 which closely matches the URAT
magnitude range.  URAT recovers 282 (=94.9\%) of these 297 TGAS stars
with 276 (=92.9\%) having a URAT parallax $\ge$ 30 mas.  In
Figure~\ref{th25urat} we show a comparison of the TGAS 25 pc sample
and URAT parallax solution.

In order to investigate how reliable the URAT parallax formal errors
are, we compared them to TGAS data.  Details are summarized in
Table~\ref{table_errT}.   A total of 31,107 stars are in common between
TGAS and our URAT south parallax solution table.  After employing basic 
cuts for number of observations, parallax errors and epoch span 13,700
stars remain in our first sample (case 1 in Table~\ref{table_errT}).
Case 2 uses the same cuts but furthermore limits the sample to stars
with a TGAS parallax of 40 mas or larger and a URAT parallax of at
least 32 mas.  
For each case, unweighted RMS values were calculated for individual
catalog parallax errors, the formal error of the parallax differences,
and the observed spread of the parallax differences.
Assuming the error estimates of the TGAS data are correct, 
both cases show that our URAT parallaxes formal errors are
underestimated by 26\%.
However, Figure~\ref{th25urat} shows that the URAT parallaxes
themselves have no obvious bias when compared to the TGAS parallaxes
(25 pc sample).

% Norbert please update fill in above about formal error investagation
% done 180107
% create figures using TGAS instead of Hipparcos: Fig.6 is already good
% Fig. 7,8 need new with final sample table (TGAS,HIP,KNOWN) DONE

In the process of obtaining a clean new discovery sample for this
paper as described in $\S$~\ref{sec:NTPWPS}, we recover 623 stars (not
in Hipparcos or TGAS) having a previously published parallax $\ge$ 40
mas.  Of these recovered parallax stars, 64 have a published parallax
$\ge$ 100 mas.  In Figure~\ref{known_25}, we show the comparison
between the 623 known stars with published trigonometric parallaxes
$\ge$ 40 mas and the URAT results.  When searching for known parallax
stars, a varying radius was used with a maximal search radius of 60
arcseconds to allow match also with high proper motion targets.  The
closest star recovered from the URAT south epoch data is GJ 1061 with
a URAT parallax of 276.5$\pm$3.7 mas or 3.6 pc and a parallax of
270.86$\pm$1.29 or 3.69 pc reported in \citep{2014AJ....148...91L}.

In Figures~\ref{pie_ep_25} and \ref{pie_nu_25}, we show the relationship
between the URAT parallax error with the epoch span of
observation and number of observations, respectively for the 1764 new
plus known parallaxes from the URAT south data.  These plots
indicate as expected that the parallax error drops with more
observations over a larger period of time.  While the URAT solutions
would benefit from this extra observing, a majority of the URAT
parallax solutions have a parallax error less than 10 mas with our
initial cuts.

%%%%%%%%%%%%%%%%%%%%%%%%%%%%%%%%%%%%%%%%%%%%%%%%%%%%%%%%%%%%%%%%%%%%%%%%%%%%%%
\subsection {New Trigonometric Parallaxes Without Prior Selection}
\label{sec:NTPWPS}
%%%%%%%%%%%%%%%%%%%%%%%%%%%%%%%%%%%%%%%%%%%%%%%%%%%%%%%%%%%%%%%%%%%%%%%%%%%%%%

We ran the entire URAT south central images observation data through 
the parallax pipeline which gave us 24.7 million parallax solutions.
The initial cuts
applied to the epoch data are summarized in Table~\ref{table_cuts1}.
Using the investigation described in $\S$~\ref{sec:comparison} we
adopted a more stringent set of additional cuts shown in
Table~\ref{table_cuts2} in order to get a more manageable cleaner
sample.  After applying these cuts we are now left with 5556 new
candidate nearby stars.  We then removed the 701 Hipparcos+TGAS stars
leaving us with a list of 4855 candidate nearby stars.  This sample
was then matched with SIMBAD, Vizier online catalogs, RECONS 25 pc
database and other published papers to remove 625 stars having a
previous published parallax, leaving us with 4230 nearby star
candidates.

While this is a more reasonable sample to investigate, we do not take
provisions for close doubles in the URAT reduction process which may
lead to many of these new candidates likely still have an erroneous
parallax solution.  A visual inspection of the parallax residual plots
for the entire 4230 candidate nearby star sample was performed leaving
us with 1229 nearby star candidates with high quality residuals.  In
Figures~\ref{2703112} and \ref{5338238}, we show an example of a high
and low quality fit solution respectively.

We then matched all stars with 2MASS and APASS using a 10 arcsecond
match radius to add J,H,K,B,V,g,r,i photometry data, so we could plot
an HR diagrams for the 1229 nearby star candidates.

%(Figure~\ref{upcsHRDb} and \ref{upcsCCb}).  In both plots the black
%dots represent known stars in the RECONS database, red dots are URAT
%stars falling in line with known main sequence stars and yellow dots
%are URAT stars indicating something might be wrong with the parallax
%or photometry.  

We then checked all stars not on the main sequence by eye using the
Aladin sky atlas tool to pull up real sky images.  We find 26 stars
with incorrect APASS magnitudes due to the large 10 arcsecond match
radius.  For these stars, no other APASS magnitude was available so we
removed the mis-match magnitudes but they were left on the list as
good parallax stars.  We then decided to do a visual inspection of sky
images for the remaining sample using the Aladin sky atlas tool.  By
doing this we remove 313 stars that might have erroneous parallax
solutions due to near-neighbor contamination.  In
Figure~\ref{upcsHRDa}, we show the HR diagram for the remaining 916
nearby stars.  Using these HR diagrams we can identify 6 possible
White Dwarf (WD) stars.  We also identify 5 possible young stars and
46 possible subdwarf stars.  These stars are flagged in the table with
more information given in $\S$~\ref{sec:Dis}.

During the visual inspection of the nearby star sample we find 19
Common Proper Motion (CPM) and parallax pairs.  Of these 12 pairs did
not have a double star designation in the SIMBAD database.  All 19
pairs are presented in Table~\ref{table_cpms} where we give the names,
parallaxes, proper motions, separation and position angle of each pair
along with notes.  For the 12 new double stars we picked the component
with the brightest V magnitude to be the component ``A''.  If a V
magnitude was not present for any of the stars in the pair we used
2MASS J magnitudes instead.  In Figure~\ref{cpms}, we show the
relationship between the parallax and proper motion of component ``A''
and component ``B'' of the CPM pairs.

The URAT north and URAT south have an overlapping region between
$+26^{\circ}$ and $-12^{\circ}$ declination. We ran a match of the 916
new URAT south 25 pc sample with the entire revised 729 URAT north 25
pc sample discussed below resulting in 67 matches.  In
Figure~\ref{overlap}, we show the comparison of the 67 stars in common
between the URAT north and south data.  Error bars have been added for
the URAT north data which on average has higher reported errors than
the south data.  While these stars are technically previously found,
we include them in both tables here because they were found again
using different URAT data.  All of these stars have been flagged in
the tables.

Of these 916 new nearby star candidates, 5 have a URAT parallax $\ge$
100 mas.  In Table~\ref{table_good}, we present details for the 916
new URAT south nearby stars having an absolute parallax $\ge$ 40 mas
(sorted by parallax).  These stars were found to have no previously
published trigonometric parallax.  We include the names, RA and Dec
coordinates (ICRS) epoch J2015.5: derived from the mean URAT position at
mean epoch along with the URAT proper motions), URAT magnitude,
estimated spectral type, epoch span, number of observations, number of
observations rejected, absolute parallax, parallax error, parallax
correction, proper motion with associated errors along with notes.
For the 151 entries in the URAT south sample having no previously
reported identification in SIMBAD we have given a USNO Proper Motion
(UPM) name.

The URAT magnitude distribution of those 916 new stars within 25 pc is
shown in Figure~\ref{nUmag}.  The distribution of the number of URAT
observations for those stars is shown in Figure~\ref{nnobs}.  In
Figure~\ref{nparerr}, we show a histogram of the parallax error for the
916 stars reported in Table~\ref{table_good} which peaks around 4 mas.
We also show in Figure~\ref{totpm} a histogram of the proper motions
for the same sample which shows many of the new nearby stars have
slow proper motions (less than 400 mas/yr).

%%%%%%%%%%%%%%%%%%%%%%%%%%%%%%%%%%%%%%%%%%%%%%%%%%%%%%%%%%%%%%%%%%%%%%%%%%%%%%
\subsection {URAT north 25 pc sample with additional cuts}
\label{sec:UNSWAC}
%%%%%%%%%%%%%%%%%%%%%%%%%%%%%%%%%%%%%%%%%%%%%%%%%%%%%%%%%%%%%%%%%%%%%%%%%%%%%%

For completeness and quality control we also ran the URAT north 25 pc parallax
results through the same stringent cuts and visual inspections we used
here for the URAT south discoveries.  This includes cutting all stars
with a parallax error greater than 14 mas, visually inspecting all
parallax fits and visually inspecting all sky images. After applying
these cuts we remove 368 stars from the URAT north 25 pc sample
leaving us with 729 having a quality fit and URAT parallax error less
than or equal to 14 mas.  All 729 URAT north results are presented in
Table~\ref{table_north}.

During the visual inspection of the URAT north nearby star sample we
find 4 CPM and parallax pairs.  Of these 2 pairs did not have a double
star designation in the SIMBAD database.  All 4 pairs are presented in
Table~\ref{table_cpmn} where we give the names, parallaxes, proper
motions, separation and position angle of each pair along with notes.
For the 2 new double stars we picked the component with the brightest
V magnitude to be the component ``A''.  If a V magnitude was not present
for any of the stars in the pair we used 2MASS J magnitudes
instead. The URAT north 25 pc stars were then matched with TGAS
revealing 33 stars all of which have a TGAS parallax $\ge$ 30 mas.

In Figure~\ref{upcnHRD} we show an HRD for the north sample.  From
this diagram we identify 10 possible WD stars.  LSPM J0543+3637, one
of the WD candidates has not been previously identified as a WD.  We
also identify 60 possible subdwarf stars.  These stars are flagged in
the table with a spectral type of ``WD'' or ``VI'' respectively..

\section{Discussion of the Southern Sample}
\label{sec:Dis}
The HRD offers a fundamental map of stellar astronomy to
understand a star's luminosity class and effective temperature without
acquiring a spectrum. All stellar objects including bright luminosity
giants, faint evolved white dwarfs, hot OB stars and cool brown dwarfs
would be separated on the HRD because of their different structures
and masses. In this work, we utilize the URAT parallaxes and
photometry gathered from the APASS and 2MASS catalogs, we can separate
our samples into different luminosity classes, including main-sequence
dwarfs, low metallicity subdwarfs, white dwarfs and young stars.

\subsection{Red Dwarfs}
\label{sec:Disr}

We utilize the K-M spectral transition at $V-K_s=$3.7 defined in
\cite{Clements2017}. If a star has no APASS $V$ magnitude, we use the
absolute magnitude at the $K_s=$5.1, which is derived using known
nearby dwarfs, to separate K and M main sequence dwarfs.

The majority of targets (864 or 94\%) presented in Table~\ref{table_good}
are main-sequence dwarfs and they have spectral types labeled as
either ``M'' or ``K''. The large bias toward M dwarfs is expected
because 70\% of the population in the Galaxy is M dwarfs and brighter
and earlier types of dwarfs are mostly been measured by the Hipparcos
or Gaia in its first data release.  All of these red dwarfs reported
here have parallaxes greater than 40 mas shown in
Table~\ref{table_good}.  The closest star among this southern sample
is LHS 5264 ($V-K_s$=5.17, $M_{K_s}=$8.4) at 9.1 pc or 110.3$\pm$2.8
mas. It has a nearby X-ray source about 16\arcsec~away from its epoch
2000 coordinate but it has not been discussed specifically in any
literature other than several survey catalogs.

\subsection{Cool Subdwarfs}
\label{sec:Diss}

K and M subdwarfs are intrinsically fainter than their dwarf
counterparts because of their low metallicity, so they are called
``sub-dwarf'' \citep{Kuiper1939, Jao2008}. The K and M subdwarfs are
basically scattered just below the main sequence, but their
distribution merges with early K and late M dwarfs in the $V-K_{s}$ vs
$M_{K_s}$ or $V-I$ vs $M_{V}$ plot.  Besides, because of the high
galactic velocities from billions of years of galactic heating
\citep{Gizis1997}, they usually have high tangential velocities. Hence,
these two unique features of cool subdwarfs can be used to separate
them from main-sequence dwarfs. \cite{Jao2017} demonstrated using the
HRD and tangential velocity to select cool subdwarf candidates. We
will use the same dwarf-subdwaf division line and the cutoff
V$_{tan}=$200 km s$^{-1}$ given in \cite{Jao2017} to select subdwarf
candidates in this sample.

None of our southern sample exceeds the cutoff V$_{tan}$ at 200 km
s$^{-1}$. However, many of stars are below the dwarf-subdwarf division
line and their locations make them promising subdwarf candidates. All
of these subdwarf candidates are labeled with a luminosity class of
``VI'' in Table~\ref{table_good}. For example, LP 764-14 ($V-K_s=$4.96,
$M_{K_s}$=8.45), a known M3.5 type subdwarf \citep{Reid2007}, is right
on the division line and its parallax is 42.1$\pm$3.6 mas or 23.7 pc.

We expect the following few stars are extreme-subdwarfs because of
their faintness at a given color, UPM 1010-2203, UPM 1215-0254, UPM
0057-3402, LSPM J1237+0804, and LP 651-57.

Noteworthy, HD 13043B appears to be lower/fainter than a dwarf and
could be a possible late K type subdwarf. Our parallax is about
43.9$\pm$6.9 mas for this secondary star. HD 13043A is the common
proper motion primary star with a spectral type of G2V
\citep{Bidelman1985} and also has independent parallaxes in various
catalogs. The original Hipparcos catalog \citep{HIP1} has a parallax
of 27.04$\pm$0.86 and the revised Hipparcos catalog \citep{HIP2} has
27.06$\pm$0.58 mas. These two measurements seem consistent with each
other but the revised parallax has a poor fit according to the
goodness-of-fit given in \citep{HIP2}.

On the other hand, the Yale Parallax Catalog (YPC) has a weighted mean
parallax of 39.6$\pm$8.6 mas, calculated from 3 independent
parallaxes of 47.8$\pm$15, 38$\pm$13.7 and 32$\pm$16 mas. Apparently,
the parallax of the primary star in the YPC is consistent with our
measurement, but differs from the Hipparcos data.

Both the Geneva-Copenhagen survey \citep{Holmberg2009} and the PASTEL
catalogue of stellar parameters \citep{Soubiran2010} showed this G2V
dwarf has a [Fe/H]$>$0 or is a metal-rich dwarf. We would need further
spectroscopic observation of the secondary and as well as parallaxes
for both components from Gaia to confirm the subdwarf feature of HD
13043B and their associations. We tentatively assign it as a subdwarf.

\subsection{White Dwarfs}
\label{sec:Disw}

Like main sequence dwarfs, white dwarfs also form a sequence far below
the main-sequence, but mixed with different types of white dwarfs, i.e
DA, DB, DZ, etc \citep{Subasavage2017}. In this work, we identify
3 new nearby white dwarfs using their locations on the HRD and
confirm 3 previously identified white dwarfs.

{\bf GD 31 (WD 0231-054)} is a known DA white dwarf with an estimated
effective temperature of 17,470K \citep{Gianninas2011}. They also
reported its spectroscopic distance is $\approx$29 pc, but our
trigonometric parallax is 55.0$\pm$10.4 mas or $\approx$18.2 pc.

{\bf UPM 0812-3529 (WD 0810-353) and UPM 0837-5017
  (WD 0836-501)} are two newly identified white dwarfs. They have
parallaxes of 104.4$\pm$5.9 and 43.1$\pm$5.2 mas, respectively. This
makes UPM 0812-3529 the closest white dwarf we have discovered in this
survey.

{\bf EC 20173-3036 (WD 2017-306)} was first detected by the
Edinburgh-Cape Blue Object Survey as a blue object with an ultraviolet
excess \citep{Odonoghue2013}. Based on its location on the HRD, it is
also a new nearby white dwarf at $\approx$17.3 pc or 57.9$\pm$4.1 mas.

{\bf LP 873-78 (WD 2118-261)} was first identified by \cite{Ryan1989}
as a possible white dwarf using $UBVRI$ colors. Its location on the
HRD confirms it is a white dwarf.

{\bf GD 1192 (WD2333-165)} is also a spectroscopic confirmed DA white
dwarf and it has an estimated effective temperature of 13,790K
\citep{Gianninas2011}. They also reported its spectroscopic distance
is about 31 pc, but our trigonometric parallax is 41.6$\pm$4.1 mas or
$\approx$24 pc.

\subsection{Young Stars or Double Stars}
\label{sec:Disy}

Young stars are elevated above the main-sequence because they are
still contracting along the Hayashi track onto the main-sequence
\citep{Hayashi1961}. These pre-main sequence stars often have active
atmospheres and emit strong H$\alpha$ or can be detected in X-ray, but
young stars older than 600 Myrs may have saturated X-ray activity
\citep{Zuckerman2004}. So, using activity alone may not be a good
indicator for youth. Here, we use both their locations on the HRD and
atmospheric activities as good indicators to select a few possible
young stars.

An un-resolved binary can also been elevated on the HRD because of
their combined luminosities and its close separation may enhance its
atmospheric activity. Hence, the young star candidates we highlight
here would need further spectroscopic observations to ensure their
youth and high resolution imaging techniques to resolve possible
doubles.

{\bf GJ 3237} is a known M4.5 dwarf \citep{Hawley1996} with an
H$\alpha$ emission \citep{Newton2017}. It also has a X-ray source
shown in the {\it ROSAT} all-sky survey \citep{Appenzeller1998}.

{\bf ASAS J081742-8243.4} is classified as an M3.5 dwaf
\citep{Riaz2006} in the $\beta$ Pic moving group \citep{Malo2013}, and
our parallax confirms its youth and expected elevated position on the HRD.

{\bf 1RXS J114738.0+050119} has a X-ray source in the {\it ROSAT}
catalog \citep{Beuermann1999} and it is clearly elevated on the HRD.

{\bf LP 795-38} is a M4V dwarf \citep{Scholz2005} and
\cite{Gaidos2014} reported an H$\alpha$ emission during their
spectroscopic observation. \cite{Winters2015} estimated its
photometric distance of 13.3 pc and our parallax is 42.29$\pm$3.5 mas
or 23.6 pc. Its youth or unresolved binary could make its photometric
distance closer than distance calculated from the parallax.

{\bf LHS 5273} has an estimated photometric distance at 16.8 pc
\citep{Winters2015} and our parallax is 44.0$\pm$4.1 mas or 22.7
pc. The {\it ROSAT} all-sky bright source catalog has a X-ray source about
21\arcsec~away and no other stellar source but LHS 5273 is within this
radius \citep{Voges1999}. Because of its elevated location on the HRD,
a possible X-ray source, and a mis-matched photometric distance, we
expect LHS 5273 is either a young star or an unresolved double.

{\bf LHS 6419} is a M3V dwarf \citep{Newton2014}. It has no X-ray
source in the {\it ROSAT} catalog, but has H$\alpha$ emission
\citep{Newton2017}.

{\bf LP 704-15} is a system with a total of 4 stars. We measure
independent parallaxes of the wide common proper motion pair for the
primary (LP 704-15) and secondary (LP 704-14). They have parallaxes of
53.0$\pm$4.7 and 48.2$\pm$4.4 mas, respectively. The LP 704-15 itself
is a SB2 \citep{Bowler2015} and also has an M8 co-moving companion with
a separation about 0\farcs47.

LP 704-15 was initially identified as a candidate of the Argus
association by \cite{Malo2013}. However, \cite{Bowler2015} argued that
LP 704-15 may be an older dwarf because 1) its stellar activity at FUV
and NUV could be produced by the SB2, not youth, and 2) the wide
companion, LP 704-14 is lack of H$\alpha$ emission and could be older
than 4 Gyrs. Their argument is consistent with LP 704-14's location on
the HRD with $V-K_s$=4.83 and $M_{K_s}=$6.67, which is not
elevated. The elevated LP 704-15 is mainly caused by the combined
optical and NIR photometry from SB2 and M8.

\subsection{Targets with notes}
\label{sec:Disn}

{\bf 2MASS J01130536-7050378} is noted as a possible red subgiant
based on color-mag diagram by \cite{Boyer2011} in the direction of
Small Magellanic Cloud (SMC). However, it has a proper motion about
76.9 mas y$^{-1}$ from our parallax data and as well as blinking
different epochs of archival images. We think it is a foreground M
dwarf at $\approx$ 17 pc ($V-K_s$=5.91, $M_{K_s}=$8.53) in the direction
of SMC.
 
{\bf Wolf 230} with a V magnitude of 11.80 is confirmed in both the
URAT north ($\pi$ = 101.9$\pm$6.0 mas, pmra = 61.2$\pm$6.0 mas/yr,
pmdc = $-$265.8$\pm$6.0 mas/yr) and URAT south ($\pi$ = 105.4$\pm$10.4
mas, pmra = 80.4$\pm$4.8 mas/yr, pmdc = $-$275.5$\pm$4.4 mas/yr)
data to be within 10 pc with a mean distance of 9.6 pc.

%%%%%%%%%%%%%%%%%%%%%%%%%%%%%%%%%%%%%%%%%%%%%%%%%%%%%%%%%%%%%%%%%%%%%%%%%%%%%%
\section {CONCLUSIONS}
%%%%%%%%%%%%%%%%%%%%%%%%%%%%%%%%%%%%%%%%%%%%%%%%%%%%%%%%%%%%%%%%%%%%%%%%%%%%%%

With the addition of these new trigonometric parallaxes from the
southern epoch data we now have a complete all sky 25 pc URAT sample.
The URAT south data has a shorter epoch span than the north data so we
have added early epoch data to help solve for proper motions and
strengthen our parallax solutions.  Many millions of parallax
solutions were obtained from the URAT south epoch data, but only a
fraction are presented here.  Many solutions have been deemed not
significant due to the formal errors on many parallax solutions,
short epoch span, quality of the fit solution
and lack of double star fitting in the URAT astrometric reduction
process.

However, we recover 623 known nearby stars south of declination $+20^{\circ}$
confirmed with URAT trigonometric parallaxes $\ge$ 40 mas of which 64
are within 10 pc.  With this survey, we add 916 newly discovered nearby
stars to the 25 pc sample.  All new nearby stars have been discovered
without previous knowledge or selection.  

Using new stringent cuts and visual checks, we also provide here a
cleaner URAT north list of 729 stars within 25 pc.  This list was cross
checked against TGAS with 33 matches all having a URAT parallax $\ge$
30 mas and one confirmed within 10 pc.

The URAT epoch data has an overlap region between $+26^{\circ}$ and
$-12^{\circ}$ declination where we find 67 stars in common.  These
stars have been left on both Tables because they have been found using
different sets of data.  In Figure~\ref{skyplot}, we show the location
of all 1648 stars from the URAT north and south epoch data.  This plot
shows a fairly uniform sky coverage with no significant clumping.

Using the URAT parallaxes and photometric data from APASS+2MASS, we
tentatively identify 4 new WD stars, 2 new young stars and 45 new
subdwarfs.  These stars would need spectrum data for confirmation.

%%%%%%%%%%%%%%%%%%%%%%%%%%%%%%%%%%%%%%%%%%%%%%%%%%%%%%%%%%%%%%%%%%%%%%%%%%%%%%

%%%%%%%%%%%%%%%%%%%%%%%%%%%%%%%%% acknowledge %%%%%%%%%%%%%%%%%%%%%%%%%%%%%%%%

\acknowledgments

We thank the entire URAT team for making this nearby star search
possible.  Special thanks go to Todd Henry and members of the RECONS
team for help with the parallax pipeline.  This work has made use of
the SIMBAD, VizieR, and Aladin databases operated at the CDS in
Strasbourg, France.  
This work used results from the European Space Agency (ESA) space mission 
Gaia. Gaia data are being processed by the Gaia Data Processing and 
Analysis Consortium (DPAC). Funding for the DPAC is provided by 
national institutions, in particular the institutions participating 
in the Gaia Multi-Lateral Agreement (MLA). The Gaia mission website is 
https://www.cosmos.esa.int/gaia. 
The Gaia Archive website is http://archives.esac.esa.int/gaia.
We have also made use of data from 2MASS, APASS,
and the ADS service as well as the PGPLOT plotting software.
We also would like to thank the many people who contributed to our
custom astrometric software code, some of which dates way back
e.g. {\citep{ausgl}.  We benefit from the fact that Fortran code
is backwards compatible allowing us to mix recent code with
original, unchanged code written over the past 50 years.

%%%%%%%%%%%%%%%%%%%%%%%%%%%%%%%%% REFS %%%%%%%%%%%%%%%%%%%%%%%%%%%%%%%%%%%%%%%

\clearpage

%%%%%%%%%%%%%%%%%%%%%%%%%%%%%%%%%%%%%%%%%%%%%%%%%%%%%%%%%%%%%%%%%%%%%%%%%%%%%%
%%%%%%%%%%%%%%%%%%%%%%%%%%%%%%% BEGIN FIGURES %%%%%%%%%%%%%%%%%%%%%%%%%%%%%%%%
%%%%%%%%%%%%%%%%%%%%%%%%%%%%%%%%%%%%%%%%%%%%%%%%%%%%%%%%%%%%%%%%%%%%%%%%%%%%%%

%\clearpage

  \begin{figure}
  \epsscale{1.00}
  \includegraphics[angle=-90,scale=0.3]{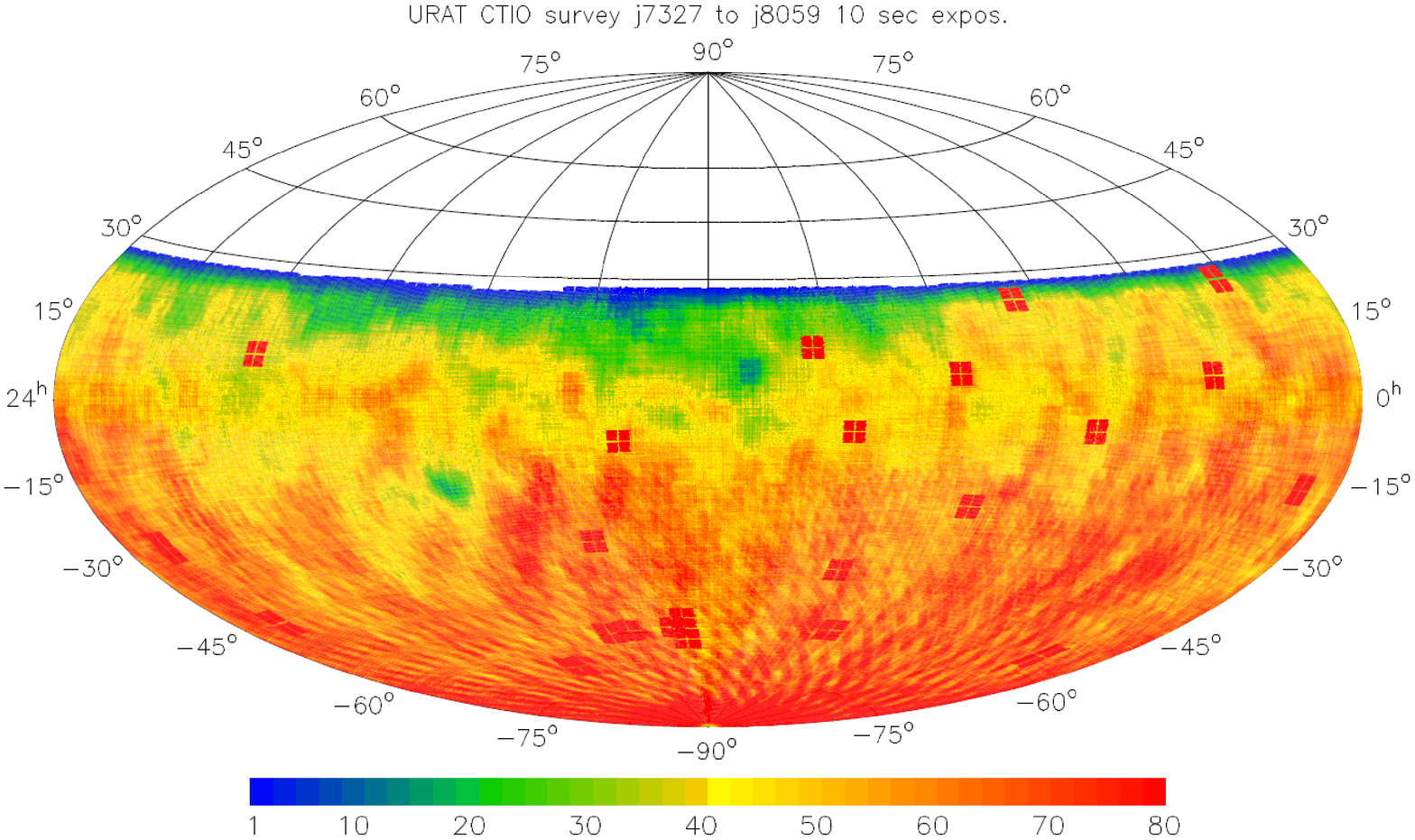}
  \caption{Sky coverage of URAT south observations used for this
    parallax investigation.  Shown here (color coded) is the number of
    10-second exposures taken for any given area on the sky.  The
    small patterns of 4 squares seen around some very bright stars is
    the URAT footprint of the 4 CCDs covering its focal plane. This
    includes observations from October 2015 (Julian Date night 7327)
    to October 2017 (Julian Date night 8059)}\label{skyurats}
  \end{figure}

 \clearpage

  \begin{figure}
  \epsscale{1.00}
  \includegraphics[angle=0,scale=0.7]{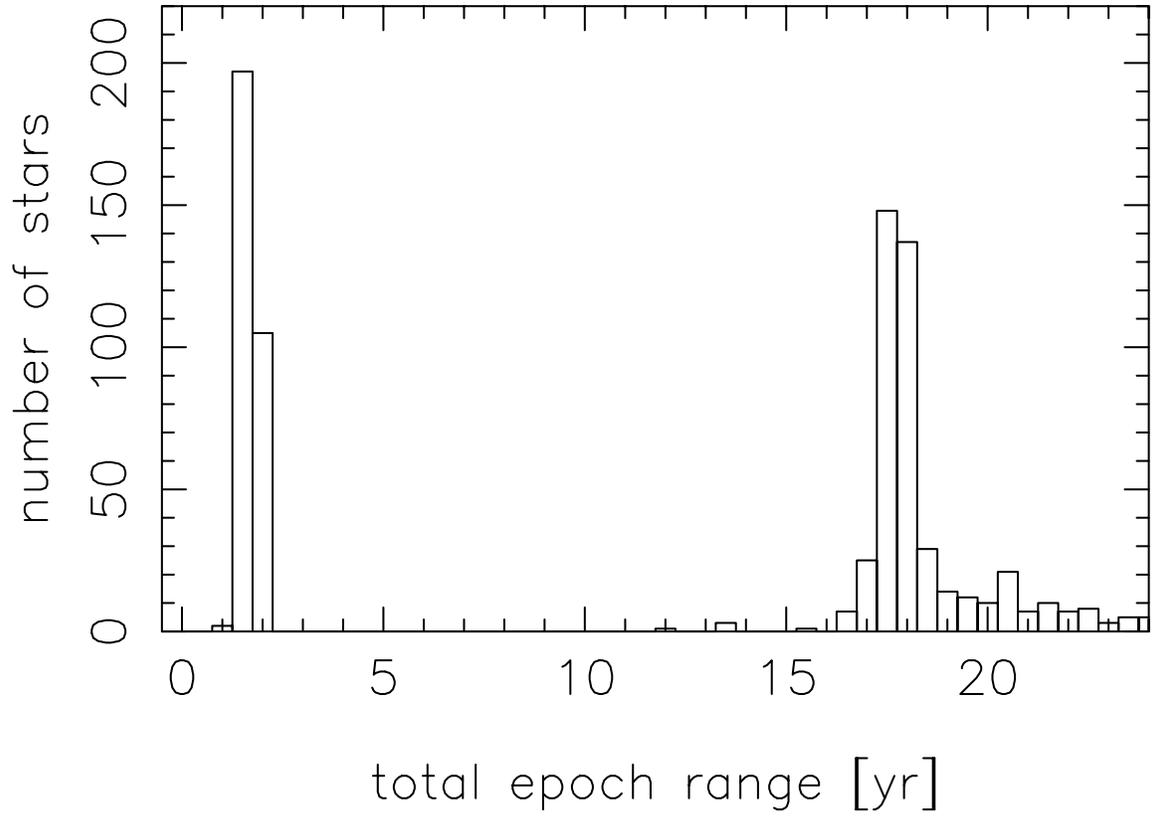}
  \caption{Histogram of the total epoch span of data used for
    our entire sample of stars of this investigation.
    The group of stars up to about 2 years of epoch span
    did not have any matched counterparts in UCAC4 or SPM4.}\label{tepoch}
  \end{figure}

 \clearpage

  \begin{figure}
  \epsscale{1.00}
  \includegraphics[angle=0,scale=0.6]{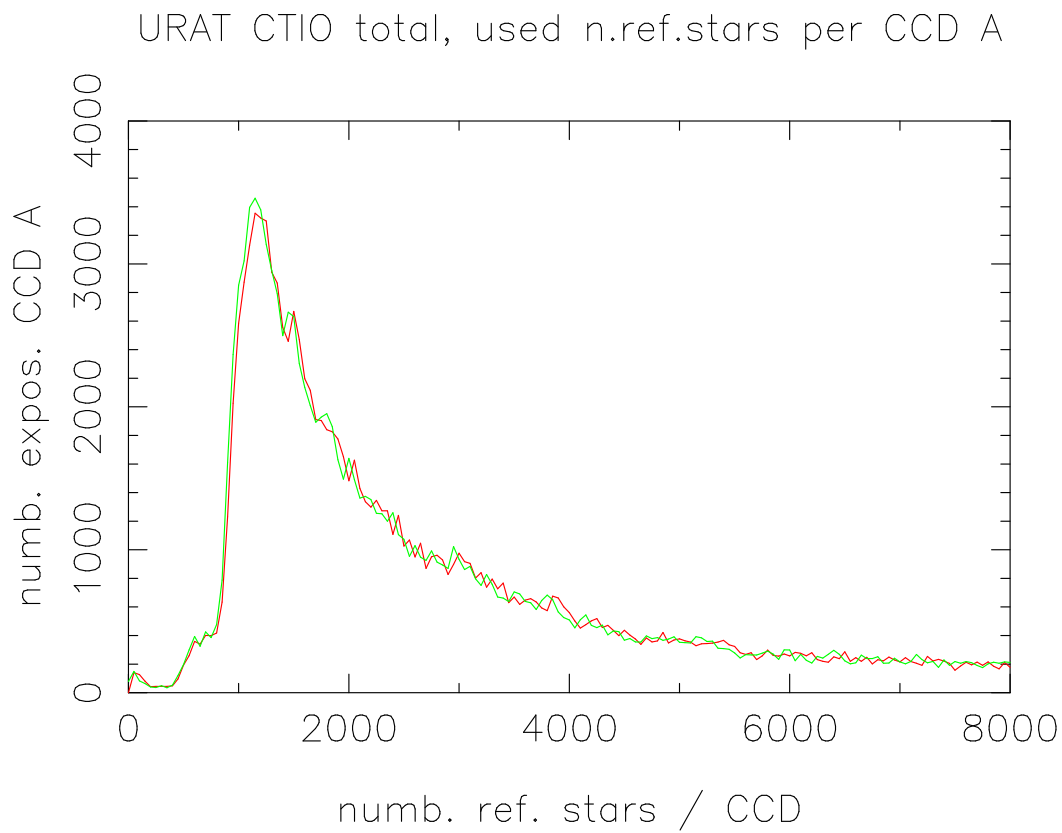} \\
  \caption{Distribution of URAT south number of reference stars on
    CCD A (results for the other CCDs are similar).
    The line slightly to the left (green) shows the number of used reference
    stars, while the other line (red) shows the total number of reference
    stars entering the astrometric solution process.
    On average, about 1 to 4\% of reference stars are eliminated in
    the reduction process (outlier rejection).}\label{asts}
  \end{figure}

 \clearpage

  \begin{figure}
  \epsscale{1.00}
  \includegraphics[angle=0,scale=0.5]{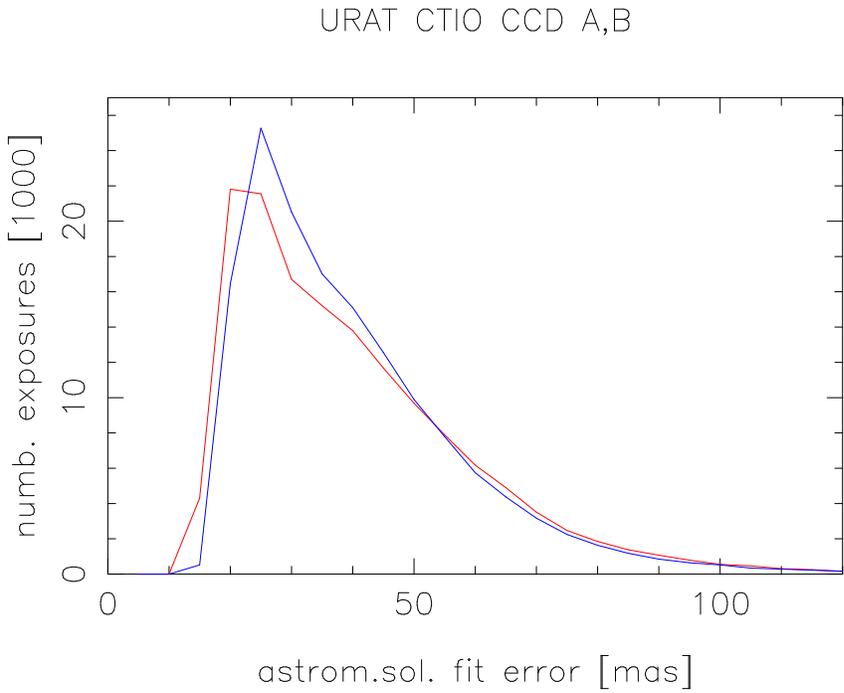} \\
  \includegraphics[angle=0,scale=0.5]{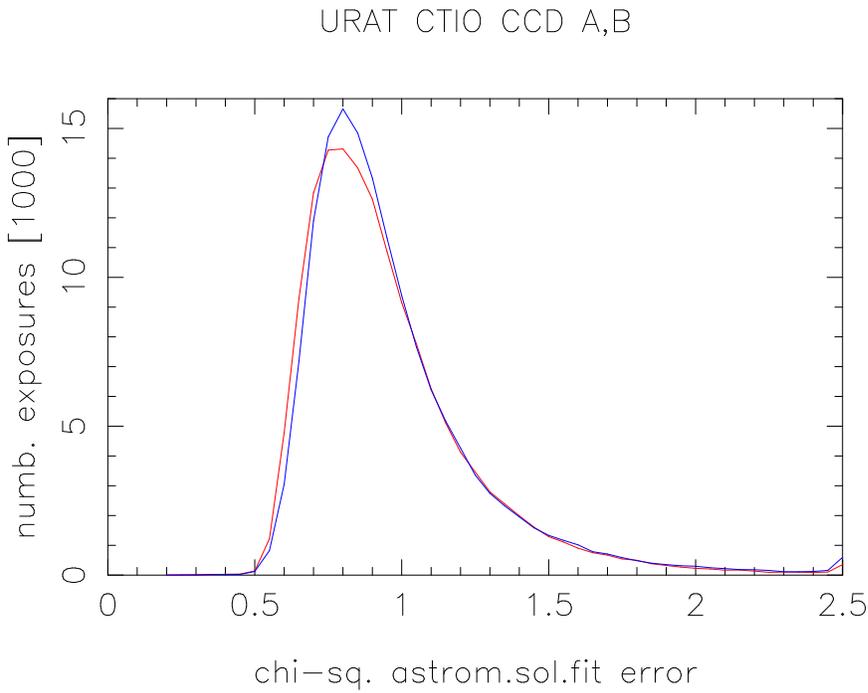}
  \caption{Distribution of URAT south astrometric solution errors
    using UCAC5 reference stars in the 8 to 15 mag range.
    Examples are shown for CCD A (red) and CCD B (blue).
    Results for the other 2 CCDs are similar.}\label{astsdist}
  \end{figure}

\clearpage

\begin{figure}
  \epsscale{1.00}
  \includegraphics[angle=0,scale=0.5]{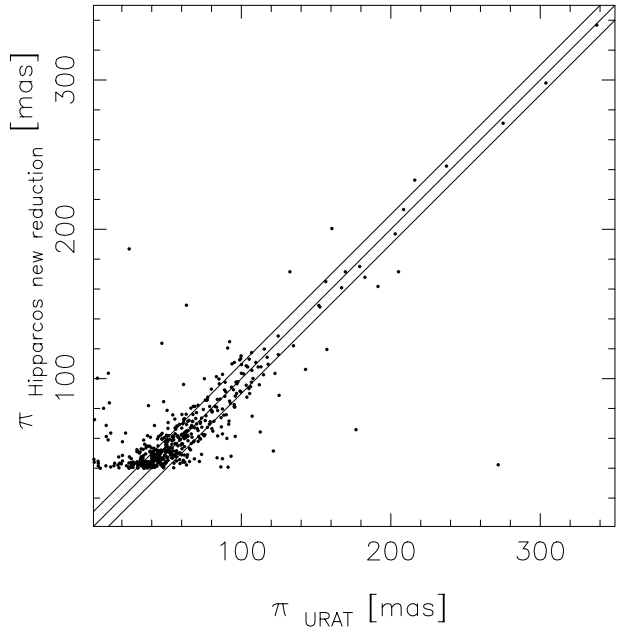} \\
  \includegraphics[angle=0,scale=0.5]{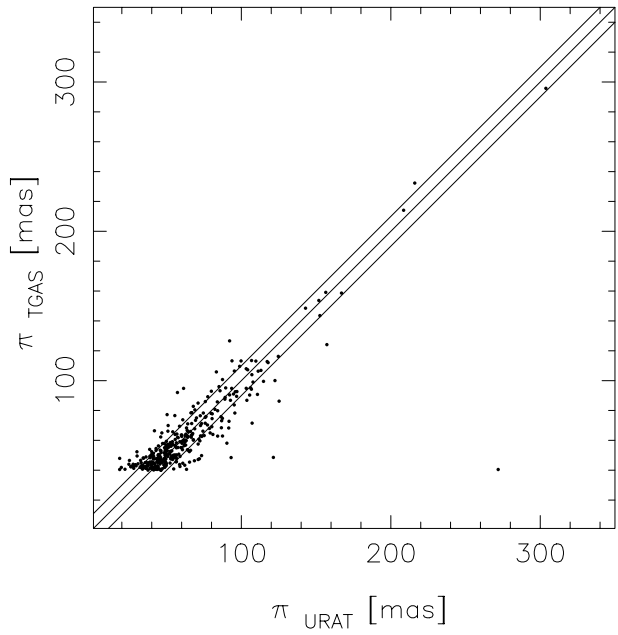}

  \caption{Parallax comparison between the URAT and the Hipparcos new
    reduction (top) and TGAS (bottom) 25 pc sample south of $\delta =
    +20^{\circ}$.  The center line represents perfect agreement while
    the outer lines show a $\pm$ 10 mas difference.}\label{th25urat}
  \end{figure}

\clearpage

  \begin{figure}
  \epsscale{1.00}
  \includegraphics[angle=0,scale=0.5]{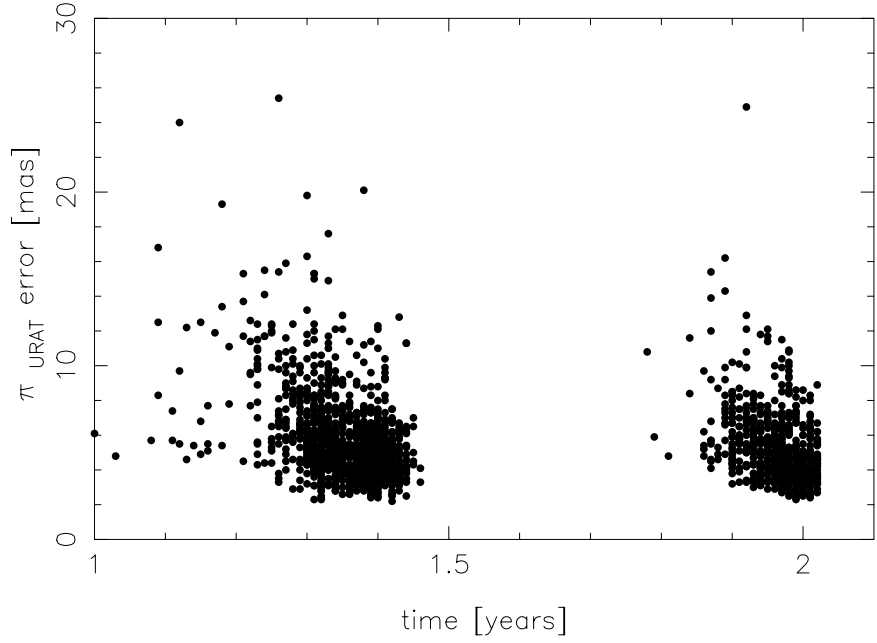} \\
  \includegraphics[angle=0,scale=0.5]{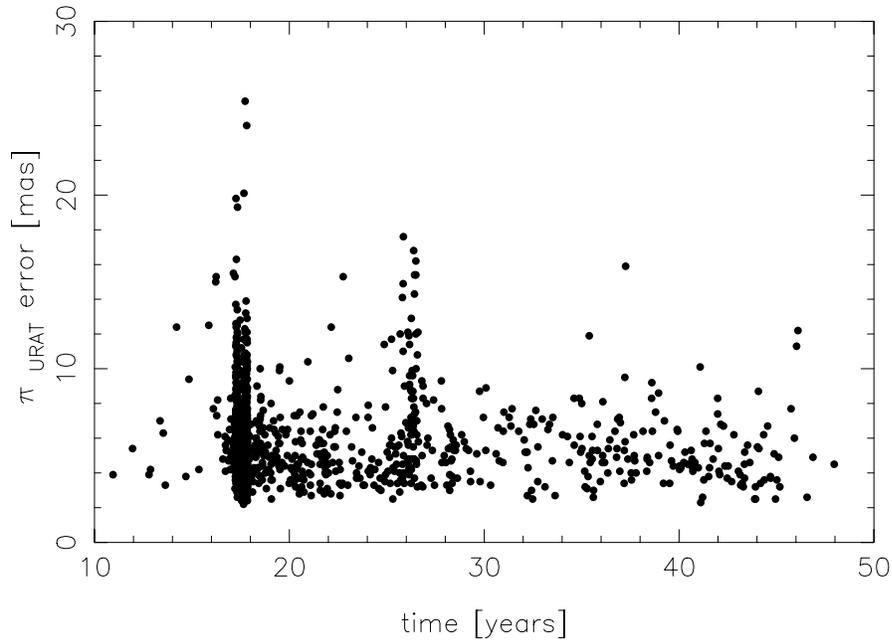}
  \caption{Relationship between URAT parallax errors and epoch span
    coverage without (top) and with (bottom) early epoch data for all
    new+known stars in this investigation with parallaxes $\ge$ 40 mas
    south of $\delta = +20^{\circ}$.}\label{pie_ep_25}
  \end{figure}

 \clearpage

  \begin{figure}
  \epsscale{1.00}
  \includegraphics[angle=0,scale=0.7]{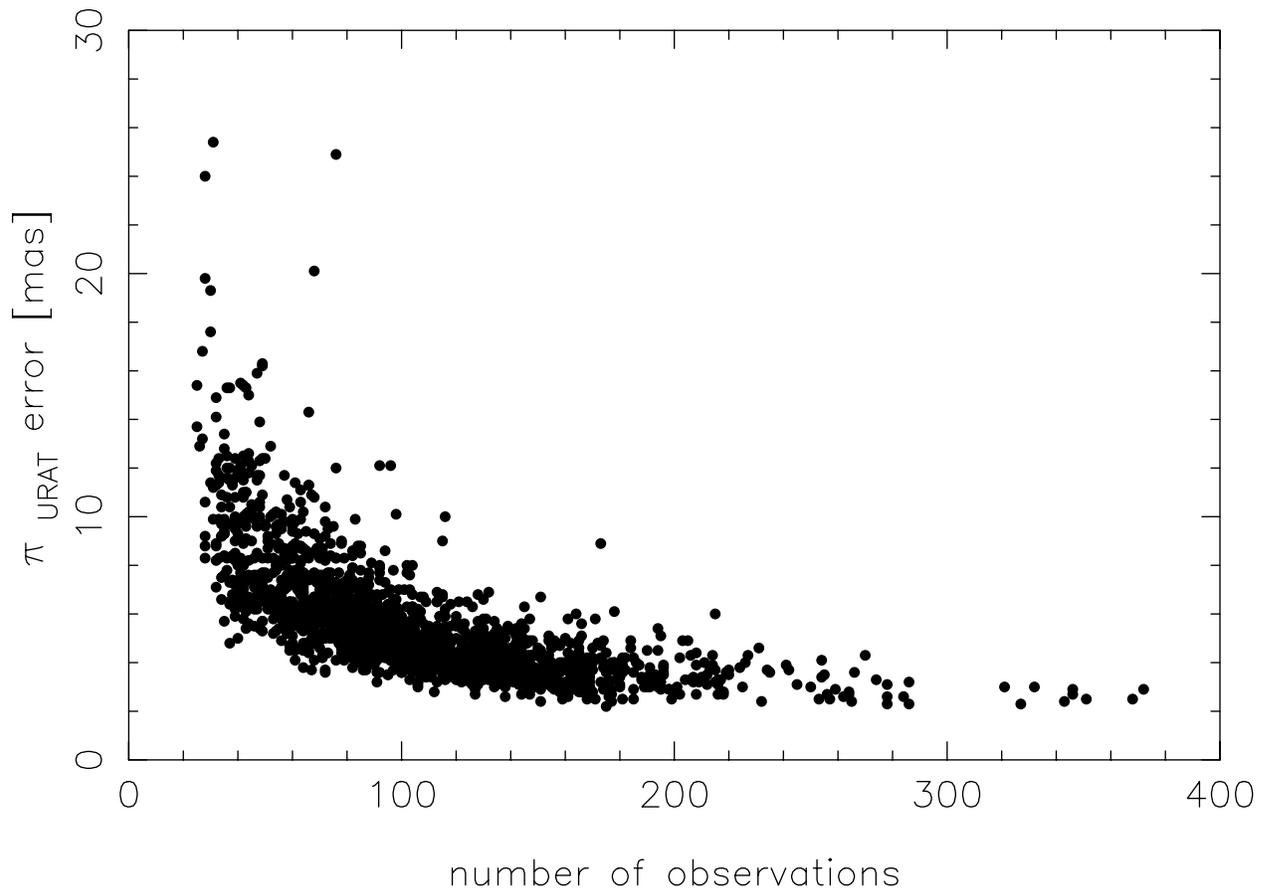}
  \caption{Relationship between URAT parallax errors and number of
    observations for all known stars with published parallaxes $\ge$
    40 mas south of $\delta = +20^{\circ}$. }\label{pie_nu_25}
  \end{figure}

 \clearpage

\begin{figure}
  \epsscale{1.00}
  \includegraphics[angle=0,scale=0.7]{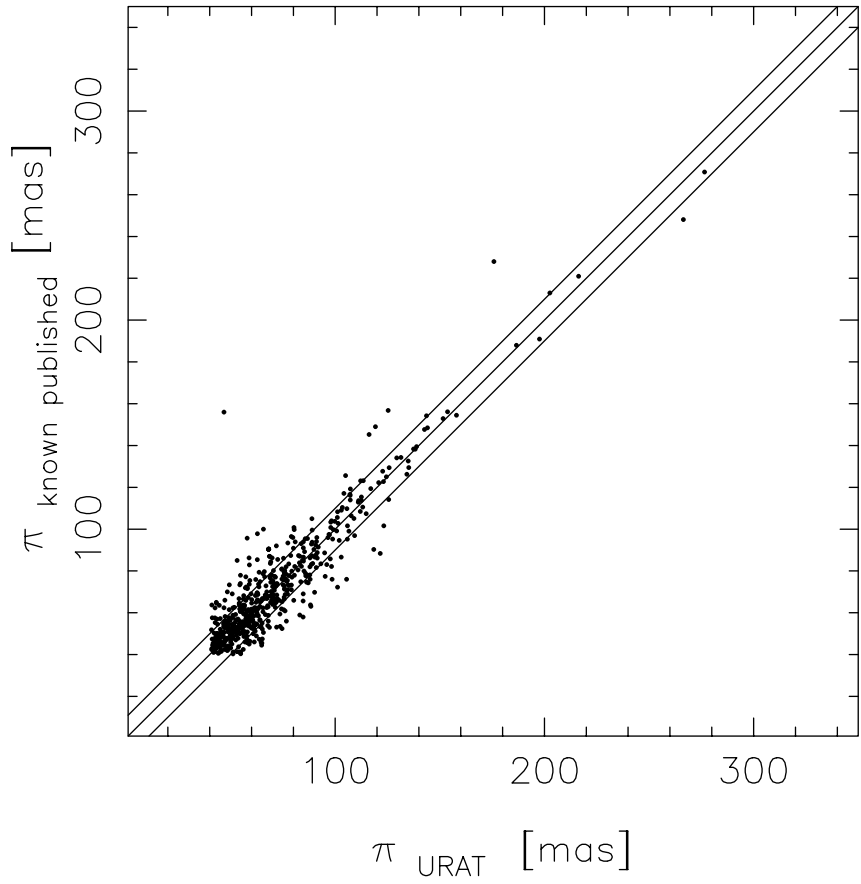}

  \caption{Parallax comparison between URAT and the 623 published 25
    pc stars south of $\delta = +25^{\circ}$ recovered in this
    investigation.  The center line represents perfect agreement while
    the outer lines show a $\pm$ 10 mas difference.}\label{known_25}
  \end{figure}

\clearpage

  \begin{figure}
  \epsscale{1.00}
  \includegraphics[angle=0,scale=0.5]{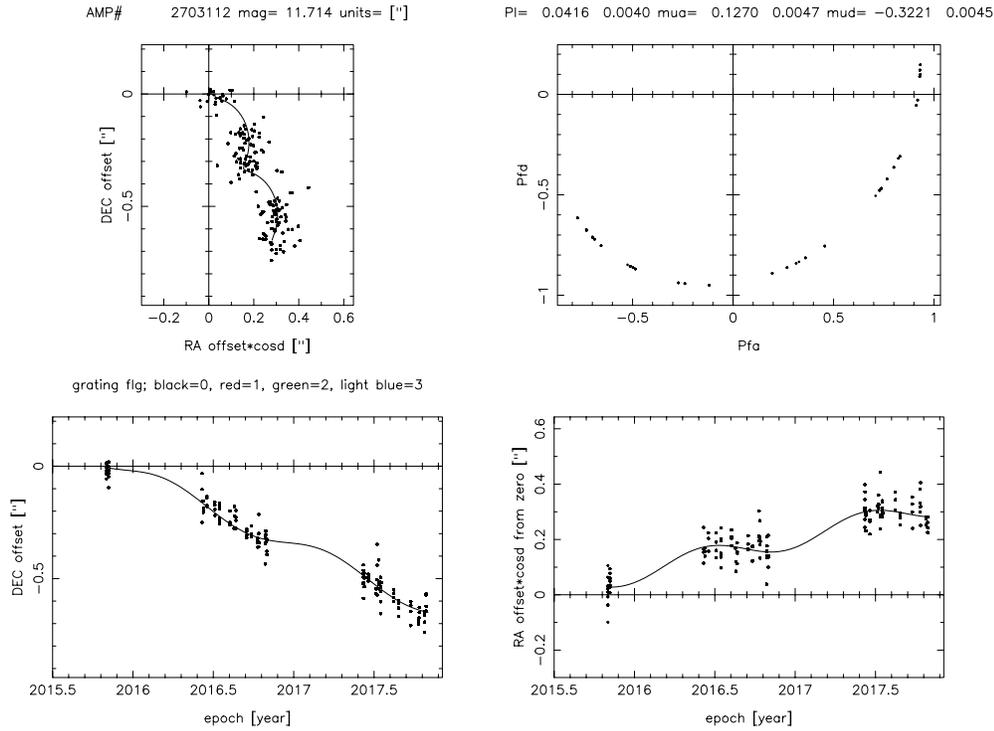}
  \caption{ Example of a good fit from our parallax pipeline.  This
    fit is for star UCAC4 080-054759 ($\pi$ = 40.6 $\pm$ 4.0 mas, 
    pmra = 127.0 $\pm$ 4.7 mas/yr, pmdc = -322.1 $\pm$ 4.5 mas/yr)
    showing in the top left
    the RA offset vs.~Dec offset [arcsecond], top right the
    parallactic ellipse, bottom left, Dec offset over time and
    bottom right RA offset over time.}\label{2703112}
  \end{figure}

\clearpage

  \begin{figure}
  \epsscale{1.00}
  \includegraphics[angle=0,scale=0.5]{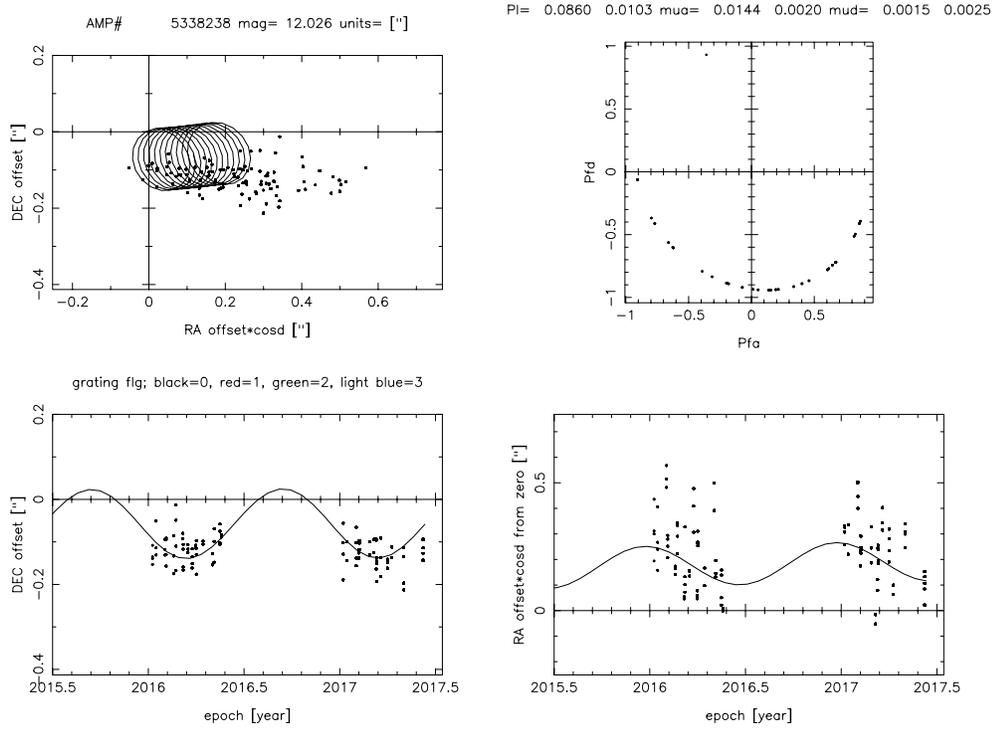}
  \caption{ Example of a bad fit from our parallax pipeline.  This
    fit is for an unknown star showing in the top left
    the RA offset vs.~Dec offset [arcsecond], top right the
    parallactic ellipse, bottom left, Dec offset over time and
    bottom right RA offset over time.}\label{5338238}
  \end{figure}

\clearpage

  \begin{figure}
  \epsscale{1.00}
  \includegraphics[angle=0,scale=0.7]{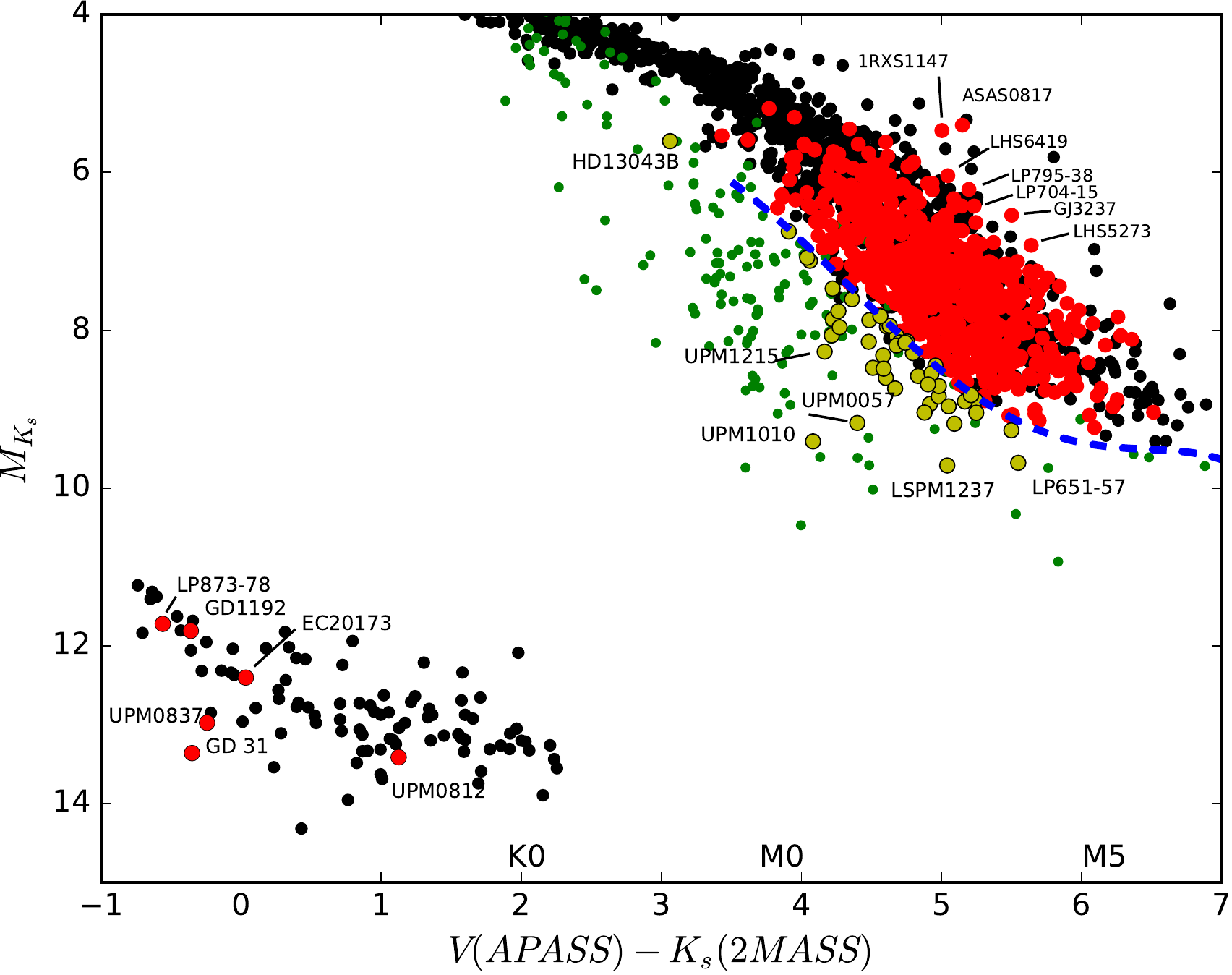}

  \caption{Stars with new parallaxes greater than 40 mas from this
    program are plotted as red circles and new subdwarf candidates are
    plotted as yellow circles. Black circles are known dwarfs and
    white dwafs within 25 pc. Green circles indicate known nearby
    subdwarfs. A blue dashed line is defined in \cite{Jao2017} to
    separate subdwafs and dwarfs. Most stars discussed in
    $\S$~\ref{sec:Dis} are labeled. We did not label a few stars
    located in the crowded region of the HRD, but their color and
    absolute magnitudes are given in $\S$~\ref{sec:Dis}. We note
    that because of limited space on this figure, a few star's
    identifiers are shorten. An approximate spectral type at a given
    color is given on the bottom of the figure.}\label{upcsHRDa}

  \end{figure}

\clearpage

  \begin{figure}
  \epsscale{1.00}
  \includegraphics[angle=0,scale=0.5]{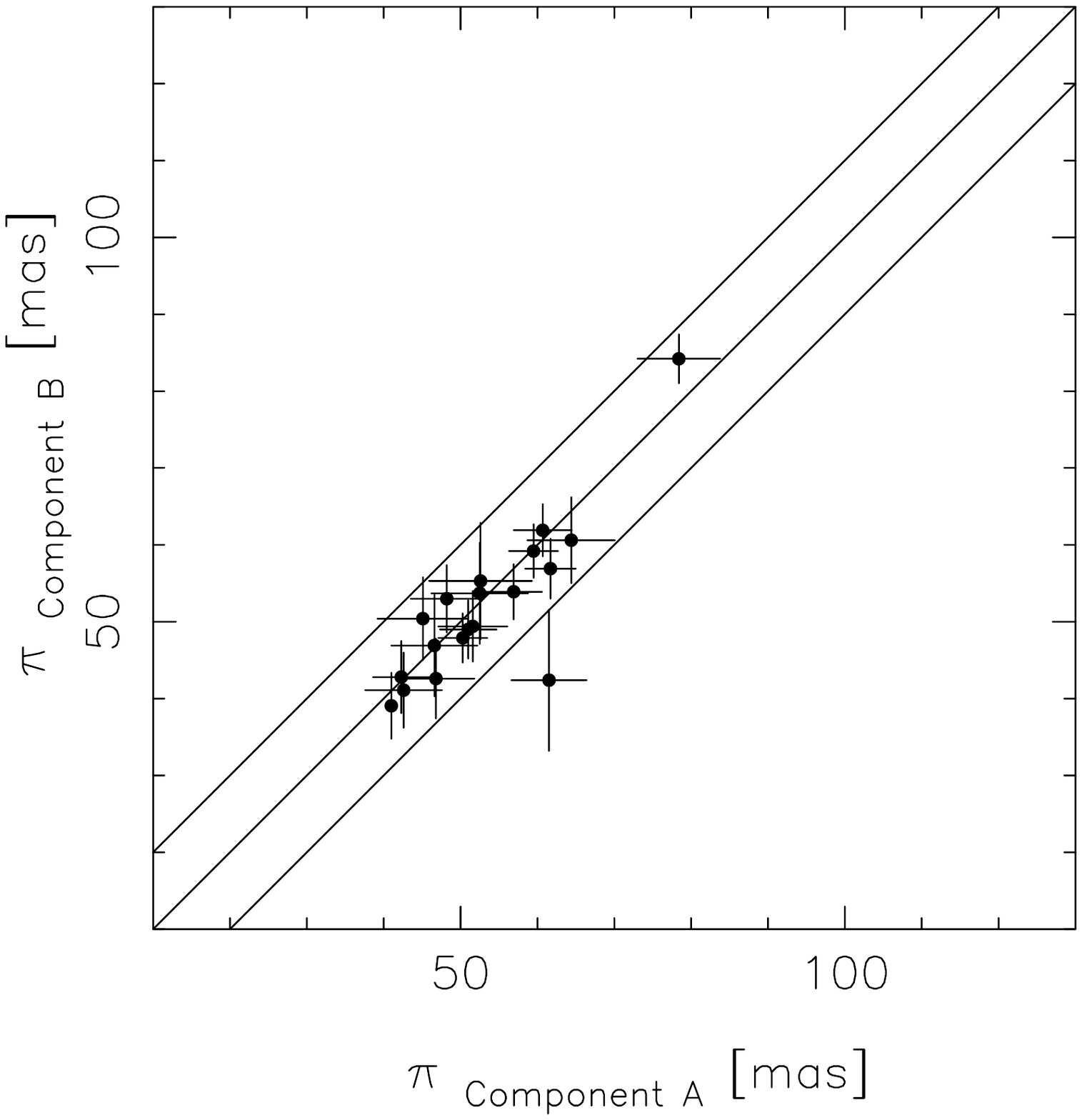} \\
  \includegraphics[angle=0,scale=0.5]{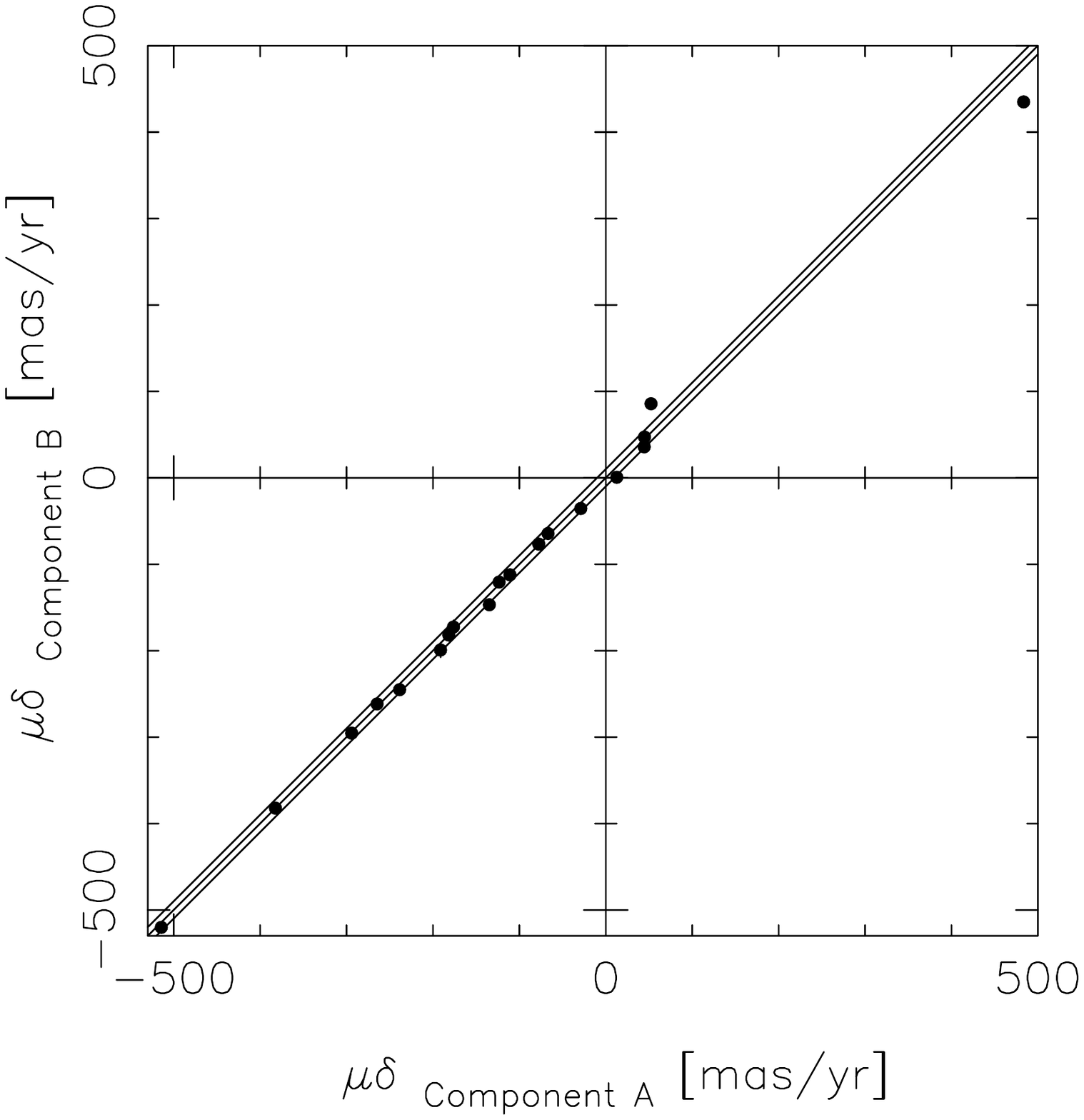}

  \caption{Relationship between URAT parallax (top) and proper motion
    in Declination (bottom) of the CPM pairs listed in
    Table~\ref{table_cpms}.  The comparison between proper motion in
    RA*cos(Dec) is intentionaly not shown because it looks the same as
    the proper motion in Declination comparison.}\label{cpms}

  \end{figure}

\clearpage

\begin{figure}
  \epsscale{1.00}
  \includegraphics[angle=0,scale=0.7]{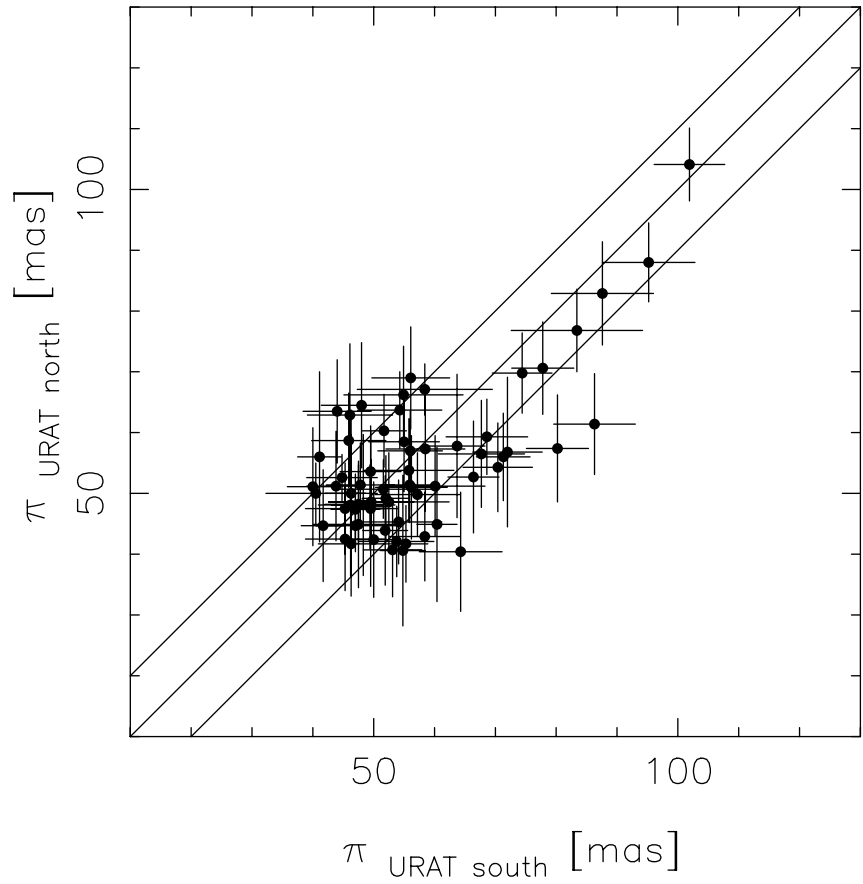}

  \caption{Parallax comparison between the 67 stars found in both the
    URAT north and URAT south data.  Error bars are shown for the URAT
    north data and the center line represents perfect agreement.}
    \label{overlap}
  \end{figure}

\clearpage

\begin{figure}
  \epsscale{1.00}
  \includegraphics[angle=0,scale=0.7]{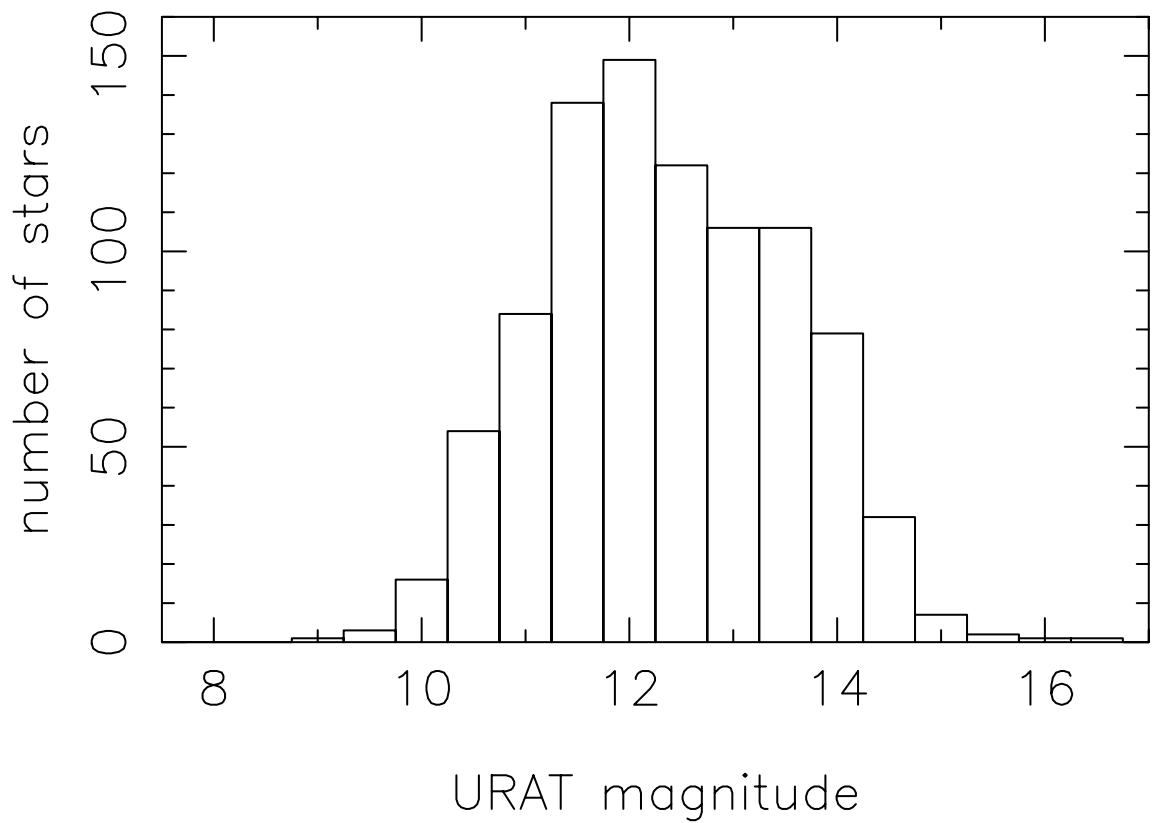}
  \caption{Distribution of URAT magnitudes (between R and I) of the
    sample of 916 newly discovered stars within 25 pc.}\label{nUmag}
  \end{figure}

\clearpage

\begin{figure}
  \epsscale{1.00}
  \includegraphics[angle=0,scale=0.7]{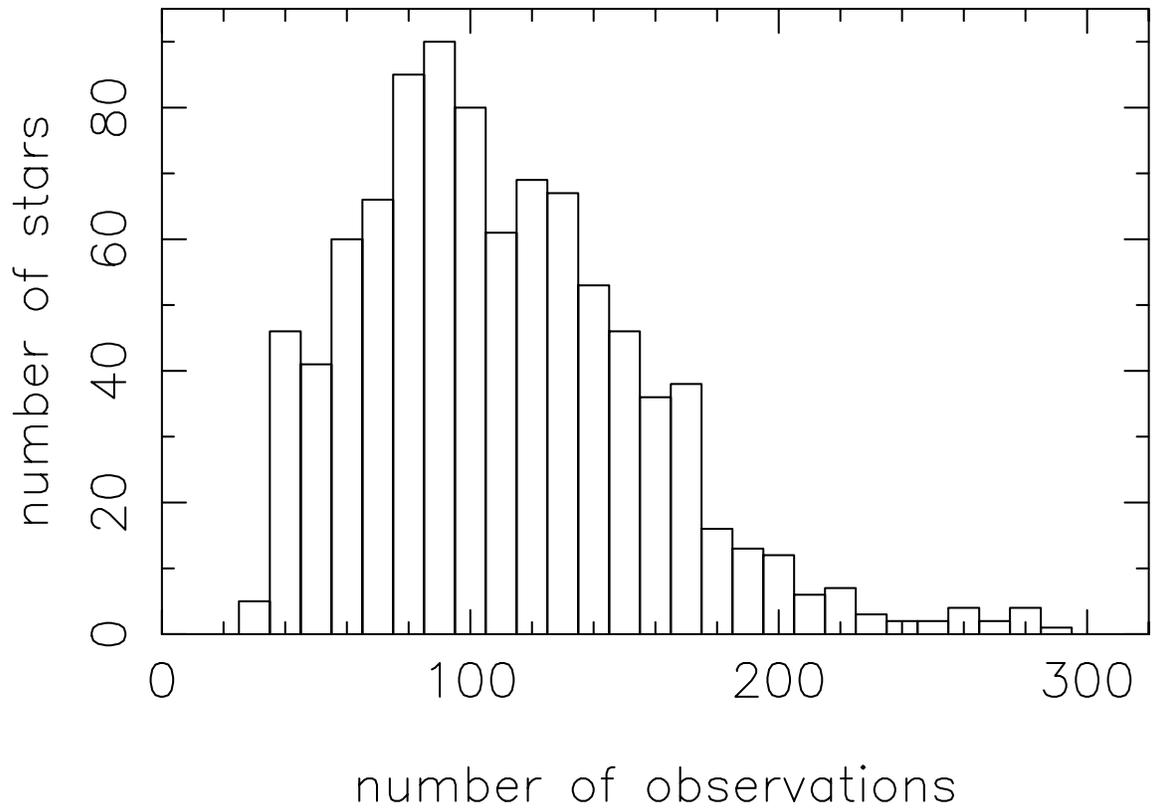}
  \caption{Distribution of number of URAT observations (exposures) of
    the sample of 916 newly discovered stars within 25 pc.}\label{nnobs}
  \end{figure}

\clearpage

\begin{figure}
  \epsscale{1.00}
  \includegraphics[angle=0,scale=0.7]{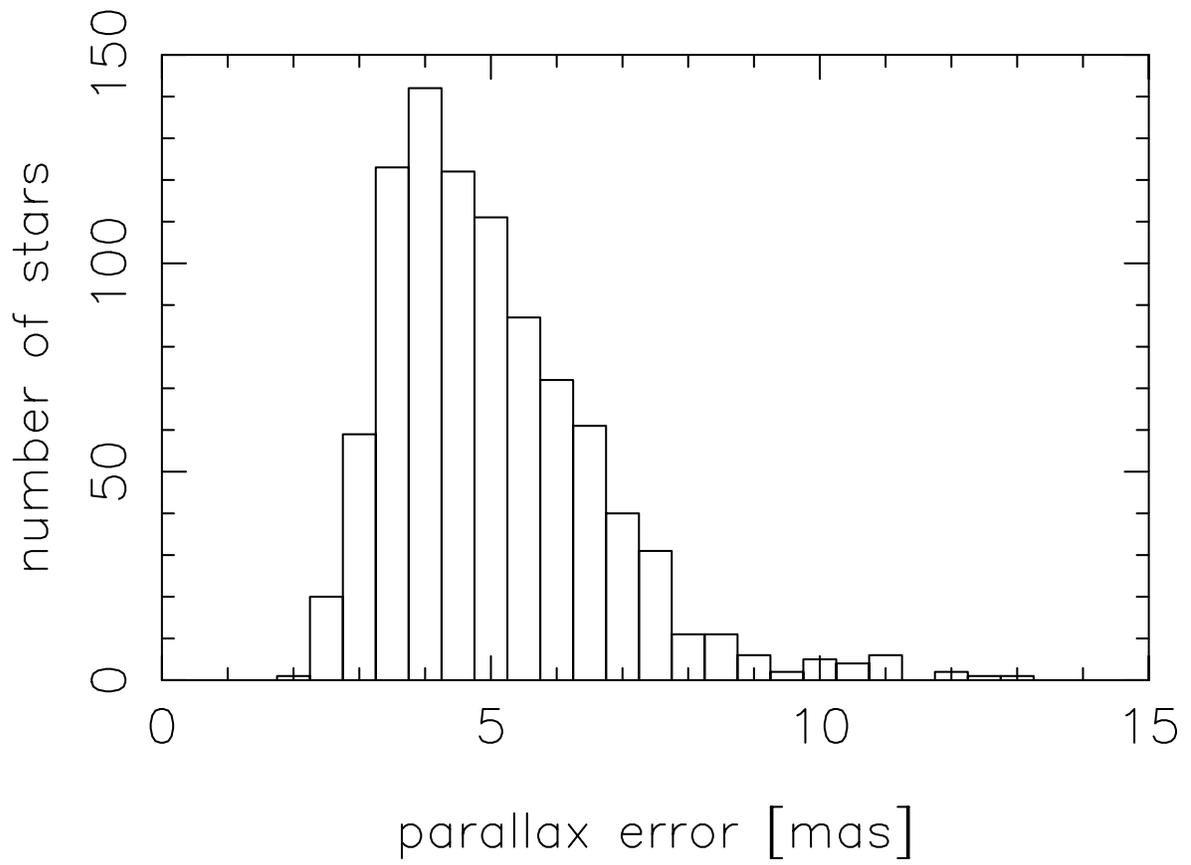}
  \caption{Distribution of URAT parallax error of the sample 
    of 916 newly discovered stars within 25 pc.}\label{nparerr}
  \end{figure}

\clearpage

\begin{figure}
  \epsscale{1.00}
  \includegraphics[angle=0,scale=0.7]{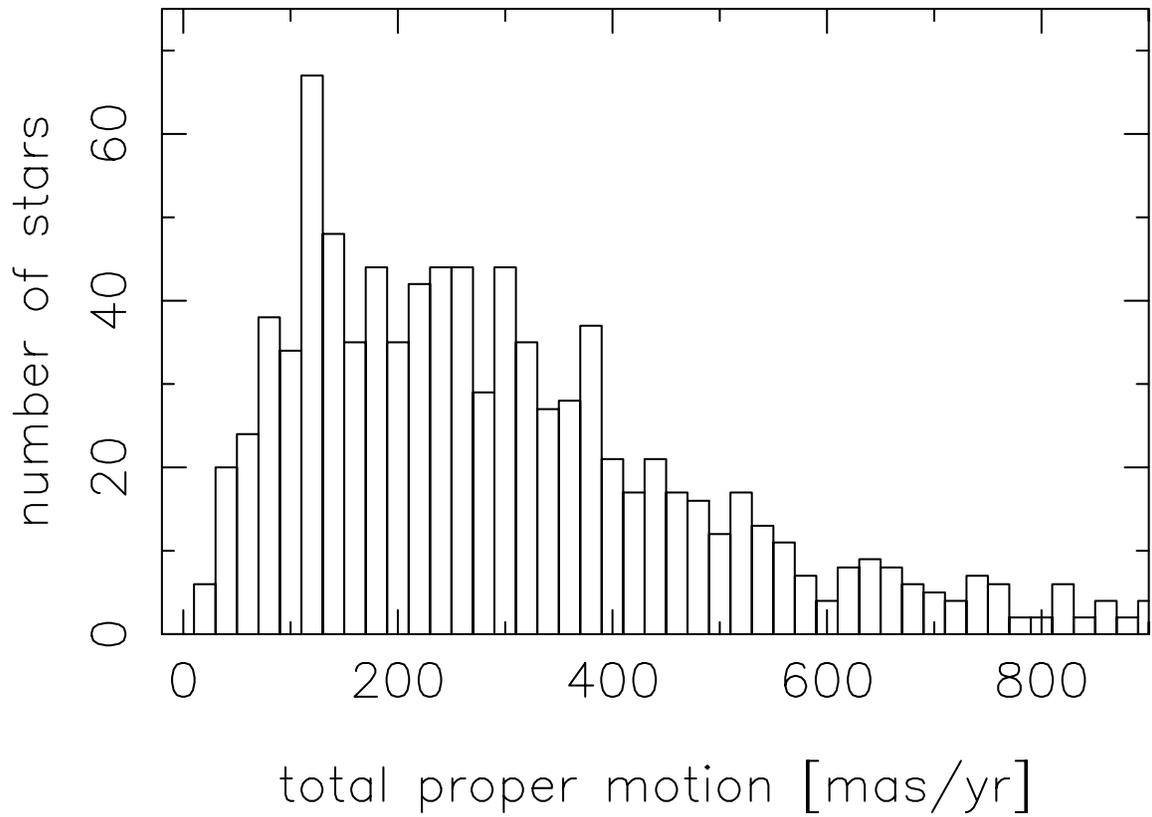}
  \caption{Distribution of URAT total proper motion of the sample 
    of 916 newly discovered stars within 25 pc.}\label{totpm}
  \end{figure}

\clearpage

  \begin{figure}
  \epsscale{1.00}
  \includegraphics[angle=0,scale=0.7]{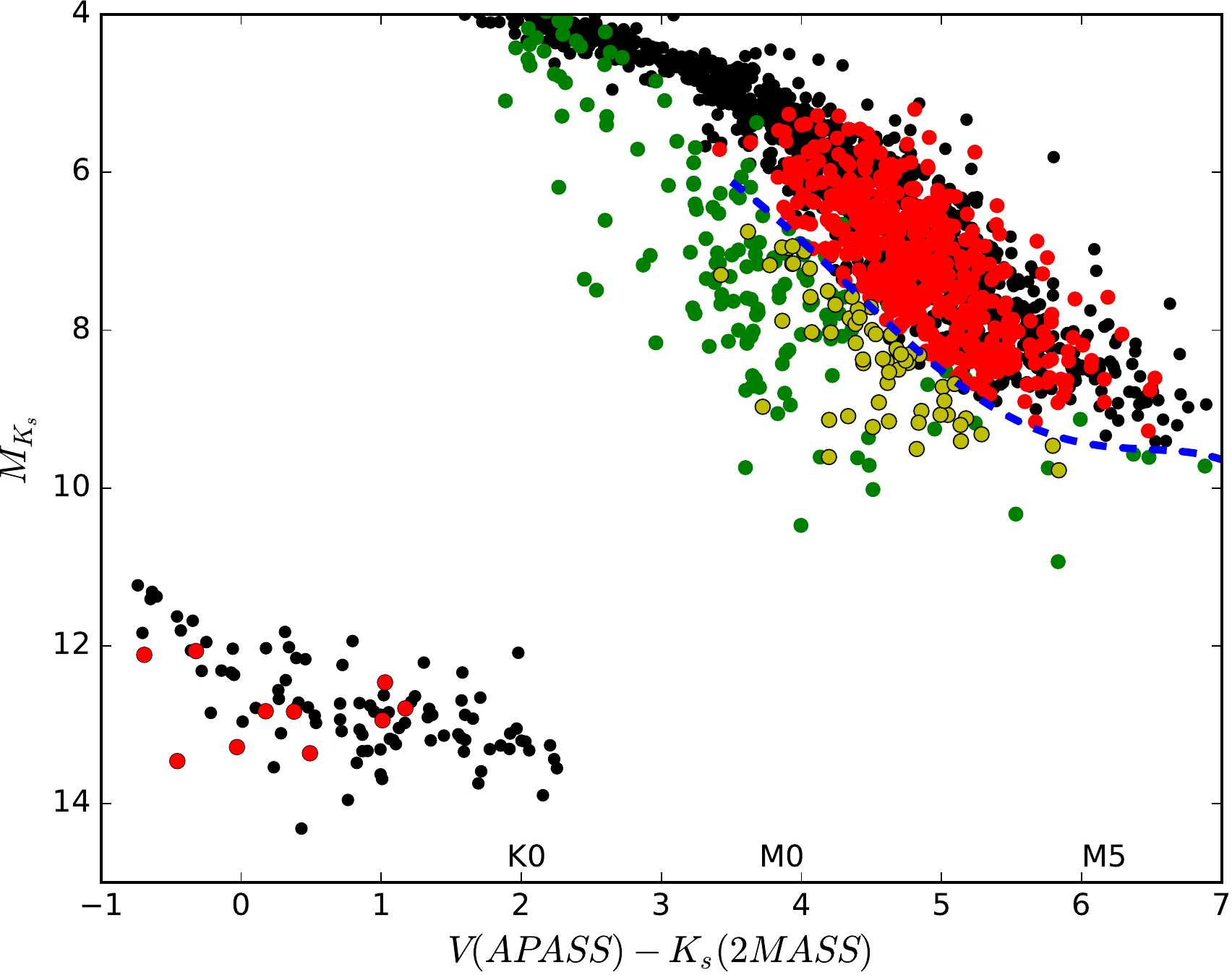}

  \caption{ The same figure as Figure~\ref{upcsHRDa}, but stars with
    parallaxes greater than 40 mas from the revised URAT north data
    are shown.}\label{upcnHRD}

  \end{figure}

 \clearpage

\begin{figure}
  \epsscale{1.00}
  \includegraphics[angle=0,scale=0.7]{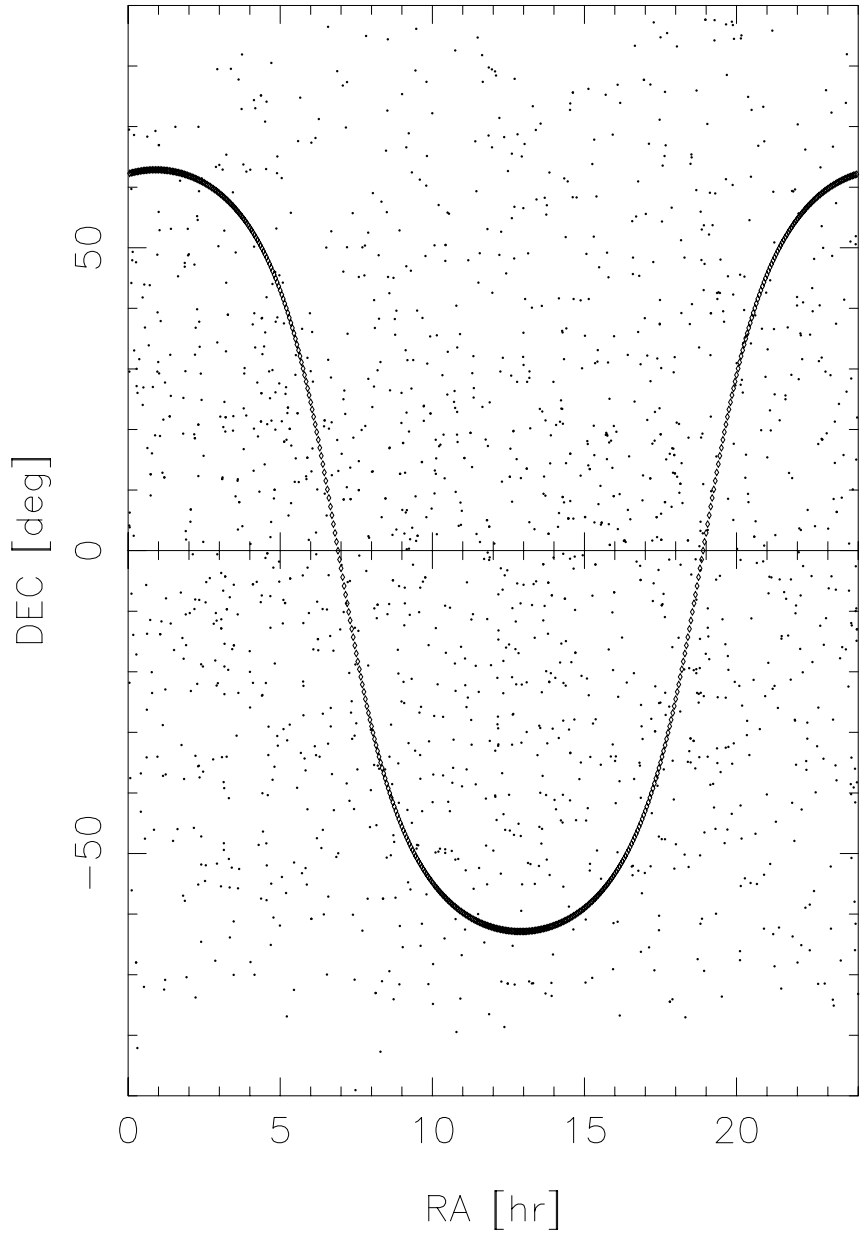}
  \caption{Distribution on the sky of all 1648 stars reported in this
    paper. The dotted line represents the galactic plane.}\label{skyplot}
  \end{figure}

%%%%%%%%%%%%%%%%%%%%%%%%%%%%%%%%%%%%%%%%%%%%%%%%%%%%%%%%%%%%%%%%%%%%%%%%%%%%%%
%%%%%%%%%%%%%%%%%%%%%%%%%%%%%% END FIGURES %%%%%%%%%%%%%%%%%%%%%%%%%%%%%%%%%%%
%%%%%%%%%%%%%%%%%%%%%%%%%%%%%%%%%%%%%%%%%%%%%%%%%%%%%%%%%%%%%%%%%%%%%%%%%%%%%%

%%%%%%%%%%%%%%%%%%%%%%%%%%%%%%%%%%%%%%%%%%%%%%%%%%%%%%%%%%%%%%%%%%%%%%%%%%%%%%
%%%%%%%%%%%%%%%%%%%%%%%%%%%%%% BEGIN TABLES %%%%%%%%%%%%%%%%%%%%%%%%%%%%%%%%%%
%%%%%%%%%%%%%%%%%%%%%%%%%%%%%%%%%%%%%%%%%%%%%%%%%%%%%%%%%%%%%%%%%%%%%%%%%%%%%%

\clearpage
\begin{deluxetable}{lrr}
\tabletypesize{}
%\rotate
%\setlength{\tabcolsep}{0.02in}

\tablecaption{Initial cuts for the URAT epoch data\label{table_cuts1}}

\tablewidth{0pt}
%tablehead{\vspace{}
\tablehead{
\colhead{CUT ITEM}            &
\colhead{LOW}                 &
\colhead{HIGH}                }

\startdata

FWHM [pixel]       & \nodata &     7.0 \\
Amplitude [ADU]    &   500   &   30000 \\
sigma RA [mas]     & \nodata &    90.0 \\
sigma DEC [mas]    & \nodata &    90.0 \\
numb. observ. used &    25   & \nodata \\
epoch span [year]  &   1.0   & \nodata \\

\enddata
\end{deluxetable}

\clearpage
\begin{deluxetable}{lrr}
\tabletypesize{}
%\rotate 
%\setlength{\tabcolsep}{0.02in}

\tablecaption{Summary of URAT formal error investigation 
  using TGAS data (see text) \label{table_errT}}

\tablewidth{0pt}
%tablehead{\vspace{}
\tablehead{
\colhead{ITEM}                   &
\colhead{case 1}                 &
\colhead{case 2}                 }

\startdata

low limit number of observations  &  30     & 30    \\
upper limit error URAT par. [mas] &  14.0   & 14.0  \\
upper limit error TGAS par. [mas] &  3.0    & 3.0   \\
lower limit epoch span [year]     &  1.4    & 1.4   \\
upper limit par.difference [mas]  &  25.0   & 25.0  \\
lower limit URAT parallax [mas]   &   0.0   & 32.0  \\
lower limit TGAS parallax [mas]   &   0.0   & 40.0  \\
\hline
total number of stars in common   & 33107   & 33107 \\
number of stars after all cuts    & 13700   &  157  \\
\hline
RMS of TGAS parallax errors [mas] &   0.34  &  0.48 \\
RMS of URAT parallax errors [mas] &   7.08  &  6.39 \\
RMS of formal error par.diff.[mas]&   7.09  &  6.40 \\
RMS of parallax difference [mas]  &   8.95  &  8.08 \\
ratio obs. RMS diff.err./formal err. & 1.26 & 1.26 \\
\hline

\enddata
\end{deluxetable}

\clearpage
\begin{deluxetable}{lrr}
\tabletypesize{}
%\rotate 
%\setlength{\tabcolsep}{0.02in}

\tablecaption{Additional cuts for the URAT epoch data\label{table_cuts2}}

\tablewidth{0pt}
%tablehead{\vspace{}
\tablehead{
\colhead{CUT ITEM}            &
\colhead{LOW}                 &
\colhead{HIGH}                }

\startdata

parallax [mas]         &   40.0  & \nodata  \\    
parallax error [mas]   & \nodata &  14.0    \\
parallax error [mas]   & \nodata & PI/4     \\
mean elongation        & \nodata & 1.25     \\
fit solution [$\chi^2$] & \nodata & 1.25     \\

\enddata
\end{deluxetable}

\clearpage
\voffset90pt{}
\begin{deluxetable}{lcccccclccccccccl}
\rotate 
\setlength{\tabcolsep}{0.02in}
\tabletypesize{\tiny} 
\tablecaption{URAT south Common Proper Motion Sytems  \label{table_cpms}}
\tablewidth{0pt}

%tablehead{\vspace{}
\tablehead{
\colhead{Primary}                        &
\colhead{$\Pi$(abs)}                     &
\colhead{$\Pi$ err}                      &
\colhead{$\mu_{\alpha}\cos\delta$}       &
\colhead{$\mu_{\alpha}\cos\delta$ err}   &
\colhead{$\mu_{\delta}$}                 &
\colhead{$\mu_{\delta}$ err}             &
\colhead{Secondary}                      &
\colhead{$\Pi$(abs)}                     &
\colhead{$\Pi$ err}                      &
\colhead{$\mu_{\alpha}\cos\delta$}       &
\colhead{$\mu_{\alpha}\cos\delta$ err}   &
\colhead{$\mu_{\delta}$}                 &
\colhead{$\mu_{\delta}$ err}             &
\colhead{Separation}                     &
\colhead{$\theta$}                       &
\colhead{notes}                          \\

\colhead{}                               &
\colhead{(mas)}                          &
\colhead{(mas)}                          &
\colhead{(mas/yr)}                       &
\colhead{(mas/yr)}                       &
\colhead{(mas/yr)}                       &
\colhead{(mas/yr)}                       &
\colhead{}                               &
\colhead{(mas)}                          &
\colhead{(mas)}                          &
\colhead{(mas/yr)}                       &
\colhead{(mas/yr)}                       &
\colhead{(mas/yr)}                       &
\colhead{(mas/yr)}                       &
\colhead{($\arcsec$)}                    &
\colhead{($\degr$)}                      &
\colhead{}                               }

\startdata

2MASS J18291690-3059489 &  56.9  &  3.3  &  463.7  &   3.0  &  -261.7  &  3.0  &  2MASS J18291670-3059556  & 61.7 & 3.9 &   464.1 &  3.9 &   -264.5 & 3.9 &   7.24 & 200.33 &                     \\  
L  186-66               &  84.2  &  5.4  & -365.6  &   5.8  &   434.9  &  5.7  &  ** LDS  217              & 78.4 & 3.2 &  -372.9 &  3.7 &    483.6 & 3.7 &   8.71 & 204.31 &                     \\
UPM 1732-0901A          &  47.9  &  3.2  &  212.1  &   4.2  &   -76.7  &  3.7  &  UPM 1732-0901B           & 50.3 & 3.2 &   218.7 &  4.2 &    -77.6 & 3.8 &   7.68 & 162.58 &                     \\
UCAC4 085-014711        &  46.9  &  5.6  &   12.9  &   0.8  &  -172.6  &  0.8  &  UPM 0808-7301B           & 46.6 & 6.6 &    20.2 &  0.9 &   -176.3 & 7.8 &   9.82 &  50.54 &                     \\
UPM 1748-7427A          &  49.4  &  4.5  &   -9.3  &   0.6  &  -146.8  &  0.6  &  UPM 1748-7427B           & 51.6 & 4.6 &    -8.9 &  0.8 &   -134.7 & 1.0 &  14.55 &   5.95 &                     \\
UCAC4 294-197036        &  59.2  &  3.2  &   20.6  &   4.0  &  -382.2  &  3.8  &  UCAC4 294-197044         & 59.5 & 3.5 &    22.6 &  4.4 &   -382.2 & 4.3 &   7.50 & 118.68 &                     \\
** GWP  441             &  42.4  &  4.9  & -244.1  &   4.6  &  -199.3  &  4.2  &  WT 1361                  & 61.5 & 9.2 &  -247.6 &  8.5 &   -191.1 & 8.4 &  16.85 & 147.59 &                     \\
CPD-66  3810A           &  39.07 &  0.43 &  -43.21 &   0.35 &    85.71 &  0.39 &  CPD-66  3810B            & 41.0 & 4.3 &   -38.0 &  1.3 &     52.5 & 1.3 &  11.92 & 110.00 &   \tablenotemark{a} \\
UCAC4 382-001184        &  61.9  &  3.8  &   72.6  &   1.3  &    35.9  &  1.3  &  2MASS J01032096-1348231  & 60.7 & 3.4 &    75.6 &  1.2 &     44.5 & 1.2 &  24.84 & 188.79 &                     \\
LP  704-15              &  53.0  &  4.7  &  203.4  &   3.7  &     0.5  &  3.6  &  LP  704-14               & 48.2 & 4.4 &   201.9 &  3.4 &     12.9 & 3.4 &  19.61 & 294.27 &                     \\
UCAC4 485-002908        &  42.6  &  5.0  &   89.2  &   1.2  &  -120.5  &  1.2  &  2MASS J02033222+0648588  & 46.8 & 5.2 &    88.4 &  1.6 &   -123.3 & 1.6 & 110.80 &  58.40 &                     \\
LP  711-44              &  53.7  &  6.3  &  226.0  &   4.0  &  -112.2  &  3.8  &  LP  711-43               & 52.5 & 6.6 &   228.7 &  4.2 &   -110.8 & 4.0 &  19.61 & 114.27 &                     \\
GJ  3228 A              &  41.1  &  5.0  &  186.7  &   3.6  &    47.1  &  3.4  &  GJ  3229 B               & 42.6 & 4.9 &   196.5 &  3.6 &     44.9 & 3.3 &  16.35 & 171.05 &                     \\
UCAC3 59-14447          &  49.0  &  3.7  &  254.9  &   3.2  &  -181.9  &  3.1  &  UPM 0601-6047B           & 51.0 & 3.8 &   263.7 &  3.4 &   -181.5 & 3.3 &   6.82 &   6.04 &                     \\
2MASS J20350677+0218289 &  55.3  &  6.7  & -100.1  &   1.3  &   -35.4  &  1.3  &  UPM 0231-5432B           & 52.6 & 7.6 &  -101.0 &  1.8 &    -28.8 & 1.8 &  16.12 & 219.87 &                     \\
Wolf 1501               &  42.8  &  3.7  & -171.7  &   3.4  &  -245.0  &  3.3  &  2MASS J14415883-1649008  & 42.3 & 4.7 &  -167.1 &  3.9 &   -238.5 & 3.7 & 252.37 & 359.19 &                     \\
LP  560-27              &  60.6  &  5.7  & -371.0  &   1.5  &   -64.6  &  1.4  &  LP  560-26               & 64.4 & 5.6 &  -372.6 &  1.4 &    -66.7 & 1.4 &  60.86 & 255.20 &                     \\
LP  920-40              &  53.9  &  3.7  &  -39.5  &   4.8  &  -520.1  &  4.5  &  UCAC4 301-127424         & 56.9 & 3.6 &   -33.0 &  4.7 &   -514.3 & 4.4 &  11.87 & 316.02 &                     \\
LP  719-38              &  50.4  &  5.9  &  201.6  &   3.8  &  -295.5  &  3.8  &  LP  719-37               & 45.1 & 5.4 &   207.9 &  5.3 &   -293.9 & 5.1 &   6.13 &  11.69 &                     \\ 

\enddata

\tablenotetext{a}{CPD-66 3810A astrometric data from \cite{HIP2}}

\end{deluxetable}

\clearpage
\voffset80pt{}
\begin{deluxetable}{lrrrccrrrrcrrrrrrrrrrrrl}
\rotate 
\setlength{\tabcolsep}{0.02in}
\tablewidth{0pt}
\tabletypesize{\tiny}
\tablecaption{New Nearby Star discoveries URAT south epoch data\label{table_good}}

\tablewidth{0pt}
%tablehead{\vspace{}
\tablehead{
\colhead{name}                 &
\colhead{R.A.}                 &
\colhead{decl.}                &
\colhead{Umag}                 &
\colhead{eStype}               &
\colhead{epdiff}               &
\colhead{obs.}                 &
\colhead{rej.}                 &
\colhead{$\Pi$(abs)}           &
\colhead{$\Pi$ err}            &
\colhead{$\Pi$(corr)}          &
\colhead{PMRA}                 &
\colhead{error}                &
\colhead{PMDC}                 &
\colhead{error}                &
\colhead{J}                    &
\colhead{H}                    &
\colhead{Ks}                   &
\colhead{B}                    &
\colhead{V}                    &
\colhead{g}                    &
\colhead{r}                    &
\colhead{i}                    &
\colhead{notes}                \\

\colhead{     }                &
\colhead{[deg]}                &
\colhead{[deg]}                &
\colhead{[mag]}                &
\colhead{     }                &
\colhead{[years]}              &
\colhead{[num]}                &
\colhead{[num]}                &
\colhead{[mas]}                &
\colhead{[mas]}                &
\colhead{[mas]}                &
\colhead{[mas/y]}        &
\colhead{[mas/y]}        &
\colhead{[mas/y]}        &
\colhead{[mas/y]}        &
\colhead{mag}               &
\colhead{mag}               &
\colhead{mag}               &
\colhead{mag}               &
\colhead{mag}               &
\colhead{mag}               &
\colhead{mag}               &
\colhead{mag}               &
\colhead{     }                }

\startdata

LP  524-13                     &    0.5167384 &    4.1360973 & 12.51 & M     & 17.80 &  70 &   4 &    44.8 &     6.1 &     0.9 &   183.3 &     3.3 &  -201.4 &     3.2 &  10.402 &   9.849 &   9.407 &  15.802 &  14.080 &  14.908 &  13.488 &  12.144 &                                                                                                                   \\
UCAC2  23039765                &    0.6617774 &  -21.8091675 & 13.43 & M     & 42.14 &  53 &   0 &    47.6 &     6.8 &     1.0 &   -30.2 &     0.7 &    55.0 &     0.7 &  10.514 &   9.934 &   9.642 &  17.022 &  15.257 &  16.129 &  14.701 &  12.946 &                                                                                                                   \\
LHS  1009                      &    0.7805800 &    6.2753807 & 13.58 & M     & 17.81 &  80 &   4 &    43.9 &     5.7 &     1.1 &   229.2 &     4.1 &  -525.9 &     5.3 &  11.039 &  10.533 &  10.295 &  17.136 &  15.364 &  16.152 &  14.843 &  13.259 &                                                                                                                   \\
LSPM J0003+0350                &    0.7959536 &    3.8447208 & 13.99 & VI    & 17.81 &  69 &   1 &    58.2 &     6.4 &     1.0 &   -90.8 &     4.5 &  -292.2 &     4.2 &  11.292 &  10.723 &  10.445 &  18.091 &  15.943 &  16.662 &  15.282 &  13.626 &                                                                                                                   \\
2MASS J00042123-3733083        &    1.0900139 &  -37.5529976 & 12.78 & M     &  1.99 & 100 &   0 &    45.7 &     4.7 &     1.1 &   275.0 &     4.8 &  -147.4 &     4.6 &  10.382 &   9.867 &   9.611 &  16.224 &  14.614 &  15.329 &  14.031 &  12.519 &                                                                                                                   \\
\enddata

\tablenotetext{a} {parallaxes from both the URAT north and south}
\tablenotetext{b} {a common proper motion companion, see Table~\ref{table_cpms} for more info}
\tablenotetext{c} {new young star candidate}
\tablenotetext{d} {known young star}

\end{deluxetable}
\clearpage
\voffset80pt{}
\begin{deluxetable}{lrrrccrrrrcrrrrrrrrrrrrl}
\rotate 
\tabletypesize{\tiny}
\setlength{\tabcolsep}{0.02in}

\tablecaption{URAT north nearby star discoveries \label{table_north}}

\tablewidth{0pt}
%tablehead{\vspace{}
\tablehead{
\colhead{name}                 &
\colhead{R.A.}                 &
\colhead{decl.}                &
\colhead{Umag}                 &
\colhead{eStype}               &
\colhead{epdiff}               &
\colhead{obs.}                 &
\colhead{rej.}                 &
\colhead{$\Pi$(abs)}           &
\colhead{$\Pi$ err}            &
\colhead{$\Pi$(corr)}          &
\colhead{PMRA}                 &
\colhead{error}                &
\colhead{PMDC}                 &
\colhead{error}                &
\colhead{J}                    &
\colhead{H}                    &
\colhead{Ks}                   &
\colhead{B}                    &
\colhead{V}                    &
\colhead{g}                    &
\colhead{r}                    &
\colhead{i}                    &
\colhead{notes}                \\

\colhead{     }                &
\colhead{[deg]}                &
\colhead{[deg]}                &
\colhead{[mag]}                &
\colhead{     }                &
\colhead{[years]}              &
\colhead{[num]}                &
\colhead{[num]}                &
\colhead{[mas]}                &
\colhead{[mas]}                &
\colhead{[mas]}                &
\colhead{[mas/y]}        &
\colhead{[mas/y]}        &
\colhead{[mas/y]}        &
\colhead{[mas/y]}        &
\colhead{mag}               &
\colhead{mag}               &
\colhead{mag}               &
\colhead{mag}               &
\colhead{mag}               &
\colhead{mag}               &
\colhead{mag}               &
\colhead{mag}               &
\colhead{     }                }

\startdata

2MASS J00015110+6930084   &    0.4610473 &   69.5019700 & 15.32 & M    &  2.37 &  53 &   0 &    49.3 &     7.2 &     2.7 &  -157.1 &     5.8 &   -82.4 &     5.7 &  12.053 &  11.510 &  11.200 & \nodata & \nodata & \nodata & \nodata & \nodata &                                                                                                                   \\
UPM0002+2933              &    0.5839250 &   29.5519016 & 12.84 & M    &  2.27 &  65 &   0 &    47.8 &     6.1 &     1.3 &    11.1 &     4.0 &  -124.1 &     4.0 &  10.252 &   9.699 &   9.438 &  15.795 &  14.224 &  14.941 &  13.631 &  12.252 &                                                                                                                   \\
LP 192-32                 &    1.5662704 &   44.1745532 & 14.74 & M    &  2.21 &  49 &   1 &    53.5 &     5.8 &     1.1 &   273.7 &     4.1 &    51.6 &     4.1 &  11.587 &  11.018 &  10.754 &  17.502 & \nodata &  17.063 &  15.783 &  14.100 &                                                                                                                   \\
2MASS J00062445+5942444   &    1.6009103 &   59.7116745 & 14.74 & M    &  2.34 &  47 &   0 &    65.5 &     6.1 &     1.5 &  -107.8 &     4.9 &  -149.8 &     4.6 &  11.460 &  10.917 &  10.607 & \nodata & \nodata & \nodata & \nodata & \nodata &                                                                                                                   \\
2MASS J00080642+4757025   &    2.0259378 &   47.9506777 & 11.00 & M    &  2.25 &  94 &   1 &    53.2 &     6.0 &     1.1 &  -122.4 &     5.1 &    11.9 &     5.1 &   8.523 &   8.000 &   7.677 &  14.390 &  12.779 &  13.472 &  12.150 &  10.640 &                                                                                                                   \\
\enddata

\tablenotetext{a} {parallaxes from both the URAT north and south}
\tablenotetext{b} {a common proper motion companion, see Table~\ref{table_cpmn} for more information}

\end{deluxetable}
\clearpage
\voffset90pt{}
\begin{deluxetable}{lcccccclccccccccl}
\rotate 
\setlength{\tabcolsep}{0.02in}
\tabletypesize{\tiny} 
\tablecaption{URAT north Common Proper Motion Sytems  \label{table_cpmn}}
\tablewidth{0pt}

%tablehead{\vspace{}
\tablehead{
\colhead{Primary}                        &
\colhead{$\Pi$(abs)}                     &
\colhead{$\Pi$ err}                      &
\colhead{$\mu_{\alpha}\cos\delta$}       &
\colhead{$\mu_{\alpha}\cos\delta$ err}   &
\colhead{$\mu_{\delta}$}                 &
\colhead{$\mu_{\delta}$ err}             &
\colhead{Secondary}                      &
\colhead{$\Pi$(abs)}                     &
\colhead{$\Pi$ err}                      &
\colhead{$\mu_{\alpha}\cos\delta$}       &
\colhead{$\mu_{\alpha}\cos\delta$ err}   &
\colhead{$\mu_{\delta}$}                 &
\colhead{$\mu_{\delta}$ err}             &
\colhead{Separation}                     &
\colhead{$\theta$}                       &
\colhead{notes}                          \\

\colhead{}                               &
\colhead{(mas)}                          &
\colhead{(mas)}                          &
\colhead{(mas/yr)}                       &
\colhead{(mas/yr)}                       &
\colhead{(mas/yr)}                       &
\colhead{(mas/yr)}                       &
\colhead{}                               &
\colhead{(mas)}                          &
\colhead{(mas)}                          &
\colhead{(mas/yr)}                       &
\colhead{(mas/yr)}                       &
\colhead{(mas/yr)}                       &
\colhead{(mas/yr)}                       &
\colhead{($\arcsec$)}                    &
\colhead{($\degr$)}                      &
\colhead{}                               }

\startdata

UPM 2202+5537     & 50.9 &   4.8 &  145.0  &  3.6  &      0.1 &   3.3  &  UPM 2202+5538     & 49.5  &  4.8  & 140.1   &  3.7   &    1.2   & 3.3   & 23.38 &  354.90 &                    \\
CCDM J14372+7537A & 64.4 &   4.5 &  164.7  &  3.0  &   -145.6 &   2.9  &  CCDM J14372+7537B & 66.8  &  4.2  & 169.5   &  2.8   & -133.3   & 2.8   & 18.32 &  138.08 &                    \\
CCDM J22441+4030A & 44.0 &  12.4 &  -74.2  &  5.5  &   -116.0 &   5.6  &  CCDM J22441+4030B & 45.91 &  0.69 & -68.906 &  1.681 & -115.974 & 0.735 & 18.91 &  270.69 &  \tablenotemark{a} \\
HD 102634         & 28.5 &  0.49 & -205.36 &  0.54 &    2.63  &   0.41 &  UPM 1149-0019B    & 44.8  &  12.4 & -202.2  & 10.3   &    2.6   & 8.1   & 27.37 &  127.99 &  \tablenotemark{b} \\

\enddata

\tablenotetext{a}{CCDM J22441+4030B astrometric data from  \cite{gaia}}
\tablenotetext{b}{HD 102634 astrometric data from \cite{HIP2}}

\end{deluxetable}

\clearpage
%\clearpage

%%%%%%%%%%%%%%%%%%%%%%%%%%%%%%%%%%%%%%%%%%%%%%%%%%%%%%%%%%%%%%%%%%%%%%%%%%%%%%
%%%%%%%%%%%%%%%%%%%%%%%%%%%%%% END TABLES %%%%%%%%%%%%%%%%%%%%%%%%%%%%%%%%%%%%
%%%%%%%%%%%%%%%%%%%%%%%%%%%%%%%%%%%%%%%%%%%%%%%%%%%%%%%%%%%%%%%%%%%%%%%%%%%%%%

\end{document}